\documentclass[review,3p,times]{elsarticle}

\usepackage{amssymb}
\usepackage{mathrsfs}
\usepackage{color,hyperref}
\usepackage{amsmath}
\usepackage{xcolor}
\usepackage{algorithm}
\usepackage{booktabs}
\usepackage{subfig}
\usepackage{graphicx}

\definecolor{darkblue}{rgb}{0.0,0.0,0.3}
\hypersetup{colorlinks,breaklinks,
            linkcolor=darkblue,urlcolor=darkblue,
            anchorcolor=darkblue,citecolor=darkblue}

\usepackage{lineno}

\newcounter{bla}

\renewcommand{\vec}[1]{\mbox{\boldmath $ #1 $}}
\DeclareMathOperator{\sgn}{sgn}

\journal{Journal of Computational Physics}

\begin{document}

\begin{frontmatter}

\title{Deep learning of interfacial curvature: a symmetry-preserving approach for the volume of fluid method}

\author[a,b]{Asim \"Onder \corref{author}}
\author[a,c,d,e]{Philip Li-Fan Liu}

\cortext[author] {Corresponding author. \textit{E-mail address:} asim.onder@gmail.com}

\address[a]{Department of Civil and Environmental Engineering, National University of Singapore, Singapore 117576, Singapore}
\address[b]{Department of Marine Environment and Engineering, National Sun Yat-sen University, Kaohsiung 804, Taiwan}
\address[c] {School of Civil and Environmental Engineering, Cornell University, Ithaca, NY 14850, USA}
\address[d] {Institute of Hydrological and Oceanic Sciences, National Central University, Jhongli, Taoyuan, 320, Taiwan}
\address[e] {Department of Hydraulic and Ocean Engineering, National Cheng Kung University, Tainan, 70101, Taiwan}
\begin{abstract}

{\color{black}Estimation of interface curvature in surface-tension dominated flows is a remaining challenge in Volume of Fluid (VOF) methods. Data-driven methods are recently emerging as a promising alternative in this domain. They outperform conventional methods on coarser grids but diverge with grid refinement. Furthermore, unlike conventional methods, data-driven methods are sensitive to coordinate system and sign conventions, thus often fail to capture basic symmetry patterns in interfaces. The present work proposes a new data-driven strategy which conserves the symmetries in a cost-effective way and delivers consistent results over a wide range of grids. The method is based on artificial neural networks with deep multilayer perceptron (MLP) architecture which read volume fraction fields on regular grids. The anti-symmetries are preserved with no additional cost by employing a neural network model with input normalization, odd-symmetric activation functions and bias-free neurons. The symmetries are further conserved by height-function inspired rotations and averaging over several different orientations. The new symmetry-preserving MLP model is implemented into a flow solver (OpenFOAM) and tested against conventional schemes in the literature.  It shows superior performance compared to its standard counterpart and has similar accuracy and convergence properties with the state-of-the-art conventional method despite using smaller stencil. }

\end{abstract}


\end{frontmatter}


\section{Introduction}

Flows with two immiscible fluids are observed in a wide range of environmental and industrial settings. A popular method to track the motion of the interface in these multiphase flows is the Volume of Fluid (VOF) method. In VOF methods, the mass and momentum equations are solved on a fixed grid where different phases are represented by volume fractions. On small scales surface tension is important, and representing this singular force in the VOF framework is a challenge.  A major error source is the inaccurate estimation of the curvature. Despite advances in algorithms, most methods remain inconsistent, and unrealistic interface deformations are common. 


The simplest method to calculate the interface curvature is given by $\kappa=-\nabla \cdot \mathbf n$ where $\mathbf n= -\nabla  \alpha/\| \nabla \alpha \|$, where $\alpha$ is the volume fraction field. This method suffers from noisy discrete derivatives as volume fractions vary abruptly at the interface. On regular grids, a consistent and well-balanced smoothing is achieved using the height functions~\cite{Sussman2003,Cummins2005}.  The height function method has very good convergence properties~\cite{Bornia2011-rj}, but greatly struggles on coarser meshes where the number of cells  per radius of curvature is less than about 10   \cite{popinet2018numerical}. Another curvature-estimation method proposed by Cummins et al.~\cite{Cummins2005} is based on reconstructed distance function (RDF), a signed distance function from the interface.  RDF is a smoother field than $\alpha$, thus delivers more accurate derivatives. 

\begin{table}
\begin{center}
 \begin{tabular}{c ccc } 
\hline
Study &    Interface  property  &Method &Training dataset   \vspace{0.04in}   \\[0.5ex] 
\hline 
\hline 
Svyetlichnyy (2018) \cite{Svyetlichnyy2018}  &  orientation& VOF& circles  \vspace{0.04in} \\ 
Xi \textit{et al.} (2019) \cite{Qi2019-mu}   &   curvature &VOF& circles  \vspace{0.04in}\\
Patel \textit{et al.} (2019) \cite{Patel2019-an}  &  curvature &VOF&spheres \vspace{0.04in} \\
Liu \textit{et al.} (2021)\cite{Liu2021-sh}  &  curvature &particle&circles \vspace{0.04in} \\
Ataei \textit{et al.} (2021) \cite{Ataei2021}   &  position &VOF&planes \vspace{0.04in}\\
LAL.-C{\'a}rdenas et al. (2021) \cite{Larios-Cardenas2021-yf,Larios-Cardenas2021-sh}  & curvature & level set &circles, waves  \vspace{0.04in} \\
Buhendwa et al. (2022) \cite{Buhendwa2022-rt}  & \begin{tabular}{ c } position, \\ orientation \end{tabular}& level set &\begin{tabular}{ c } circles, ellipses,\\sine waves , stars \end{tabular} \vspace{0.04in}\\
Present work   & \begin{tabular}{ c } curvature \end{tabular} &VOF& \begin{tabular}{ c } circles \end{tabular}  \\

\hline
 \end{tabular} 
\end{center} 
\caption{Summary of current literature on neural networks to estimate interface characteristics in two-phase flows. }
\label{tab:literature}
\end{table}

Recently, an alternative data-driven approach has emerged to estimate interface properties. In this methodology, an input-output problem is formulated where one tries to extract from the data a mapping between the target interface property and volume fractions.  Artificial neural networks have been at the forefront of this domain. Thanks to their flexible modular structure, neural networks can capture nonlinear functional relations with high degree of complexity. Table~\ref{tab:literature} presents a short summary of the current literature.  {\color{black} Qi \textit{et al.}~\cite{Qi2019-mu} pioneered using neural networks for curvature estimation. They built a training dataset composed of circular interface segments and developed a two-dimensional (2D) model mapping the volume fractions to the non-dimensional curvature.  Patel \textit{et al.}~\cite{Patel2019-an} extended this approach to three dimensions using spherical segments. Both  Qi \textit{et al.} and Patel \textit{et al.} employed a feedforward neural network with a single hidden layer. In general, neural networks become more expressive the ``deeper" they get by stacking more layers and neurons \cite{Aggarwal2018}. Deeper architectures for curvature are explored by LAL.-C{\'a}rdenas et al. (2021) \cite{Larios-Cardenas2021-yf,Larios-Cardenas2021-sh}They also extended the training datasets to wavy configurations. }

In general, neural network models improved the predictions on coarser resolutions but lacked consistency. Results often deteriorated with grid refinement. Another overriding issue is that models are not inherently sensitive to the active symmetries of an interface. Symmetry errors can yield unphysical distortions on basic geometries. It is desirable that a neural network model delivers equivalent results when the input volume-fractions grid is rotated, reflected or sign conventions are reversed.  Invariance to these transformations is inherent to conventional methods such as the height function method. A straightforward symmetry-preserving model is proposed by Buhendwa et al. in the context of interface reconstruction~\cite{Buhendwa2022-rt}. In this model, the neural network is evaluated for the rotations and reflections of the input grid, and the results are averaged over. 7 additional evaluations were required to cover 8 considered symmetry configurations. Buhendwa et al. did not include anti-symmetric configurations where the interface is composed of convex and concave elements with the same absolute value of curvature, e.g., the peak and trough of a sine wave.  Preserving such symmetries are central to  simulation of wavy interfaces.  This requirement increases the total number of plane-symmetry configurations to 16, 8 with positive- and 8 with negative-signed curvature. Thus, 15 additional evaluations are needed if the methodology of Buhendwa et al.~\cite{Buhendwa2022-rt} is to be extended to all symmetry configurations. Such a large number demands a more efficient approach to obtain a symmetry-invariant neural network model.

{\color{black} The present work develops an efficient symmetry-preserving method to estimate the interface curvature from 2D volume fractions using neural networks. We first propose a new strategy which automatically preserves the anti-symmetries without any additional cost. This is achieved by employing a feedforward neural network with input scaling, bias-free neurons and odd-symmetric activation functions.  The output magnitude of such neural networks is independent of convexity or concavity of the interface. As a result, convex interfacial segments suffice in model building, reducing the required number of symmetry configurations to 8. These configurations are then further reduced to 4 by exploiting the information from the interface normal vector and rotating the input grid accordingly. This second approach is similar to the rotations applied in height function method. As the newly proposed method restricts the training to 4 interface configurations instead of 16, it allows a 75\% reduction in the dimension of the training dataset. 
The proposed model is implemented into the open-source library OpenFOAM~\cite{of} and is tested on various standard problems. The results are substantially improved on finer grids, and overall model performance is on par with the height function method despite using a smaller input volume-fractions grid.}

In Sec.~\ref{sec:method}, first, the relevant background information on VOF methodology and neural networks are introduced. Subsequently, the details of the new symmetry-preservation strategy are discussed. The training of neural networks are discussed in Sec.~\ref{sec:model}. Sec.~\ref{sec:testing} concerns with the  testing of the models using various two-dimensional analytical interface geometries. These tests are extended to simulation environments in Sec.~\ref{sec:simulations}, where the performance in standard benchmark cases are analysed. Finally, Sec.~\ref{sec:conclusions} presents the conclusions and future outlook.

\section{Methodology}\label{sec:method}
\subsection{Surface tension model}\label{sec:MLP}
In Volume of Fluid (VOF) methods, the mass and momentum equations are solved on a fixed grid where different phases are represented by a Heaviside function $\chi$.  This function is homogenous in both immiscible fluid phases, e.g.,  $\chi=0$ in fluid 1 and $\chi=1$ in fluid 2, with a step change from 0 to 1 at the interface. The discrete approximation to $\chi$ defines the volume fraction field, i.e., on a regular 2D Cartesian grid,
 \begin{align}
 	\alpha_{i,j}(t)&=\frac{1}{|V|}\int_{V} \chi(\mathbf x, t)\mathrm d\mathbf x, 
 	\label{eq:alpha}
 \end{align}
 {\color{black}where $|V|$ is the volume of the cell, and the subscript ($i,j$) is the cell index.}  The segment of the unknown interface in each multi-material cell ($0<\alpha_k<1$) is reconstructed with Piecewise Linear Interface Calculation (PLIC)  \cite{Rider1998,Scardovelli2003,Pilliod2004}, in which a cutplane with normal ($\mathbf n_{i,j}$) is sought to divide the cell into two subcells. 
 
 The capillary force acting at an interface is expressed by $\mathbf f^\sigma=\sigma \kappa \delta_s \mathbf n$, where $\sigma$ is the surface tension coefficient, $\kappa$ is the interface curvature and $\delta_s$ is the Dirac distribution localized at the interface. A standard model to calculate this force is the Continuum Surface Force (CSF) model proposed by Brackbill et al.~\cite{Brackbill1992}: 
\begin{equation}
	\mathbf f^{\sigma}_{i,j}=\sigma \kappa_{i,j}(\nabla  \alpha)_{i.j},
	\label{eq:CSF}
\end{equation}
where $\kappa_{i,j}$ is the interface curvature in cell ($i,j$).  We use artificial neural networks to build functional relationships between the volume fractions and $\kappa_{i,j}$. 
\subsection{Multilayer perceptron (MLP) model}\label{sec:MLP}

 An artificial neuron, or perceptron, is a mathematical function that maps a vector valued input $\hat {\mathbf x}$ to a scalar output, i.e.,
\begin{equation}
\label{eq:neuron}
	 \hat y =\phi(\mathbf w \cdot \hat {\mathbf x}+b),
\end{equation}
where $\mathbf w$ represent the weights, $b$ is the bias variable and $\phi$  is an activation function. The perceptron is the smallest building block of an artificial neural network. A layer of the neural network hosts a multitude of these units, that are fully connected to the neurons in adjacent layers. This densely interconnected system of neurons is called multilayer perceptron (MLP). The input layer in an MLP reads the database and feeds it to hidden layers where the data is processed and dumped finally to the output layer. This construct is mathematically expressed by 
 \begin{equation}
 \label{eq:MLP}
 \begin{aligned}
 	\mathbf l_1&=&\phi_1(W_1^T\hat{\mathbf x}+\mathbf b_1),\\
 	\mathbf l_2&=&\phi_2(W_2^T{\mathbf l_1}+\mathbf b_2),\\
 	&\vdots& \\
 	\mathbf l_{N_h}&=&\phi_{N_h}(W_{N_h}^T\mathbf l_{N_h-1}+\mathbf b_{N_h}),\\
 	\mathbf{\hat y}&=&\phi_o(W_{o}^T\mathbf l_{N_h}+\mathbf b_o),
 \end{aligned}
 \end{equation}
 where $W_l$ is the weights matrix in the $l$th layer. The weights $\mathbf W=\{W_1,W_2,\ldots,W_{N_h},W_o\}$ and the biases $B=\{\mathbf b_1,\mathbf b_2,\ldots,\mathbf b_{N_h},\mathbf b_o\}$ are the parameters of the MLP to be learned from data.

\begin{figure}[!t]
\begin{center}
\includegraphics[scale=0.21]{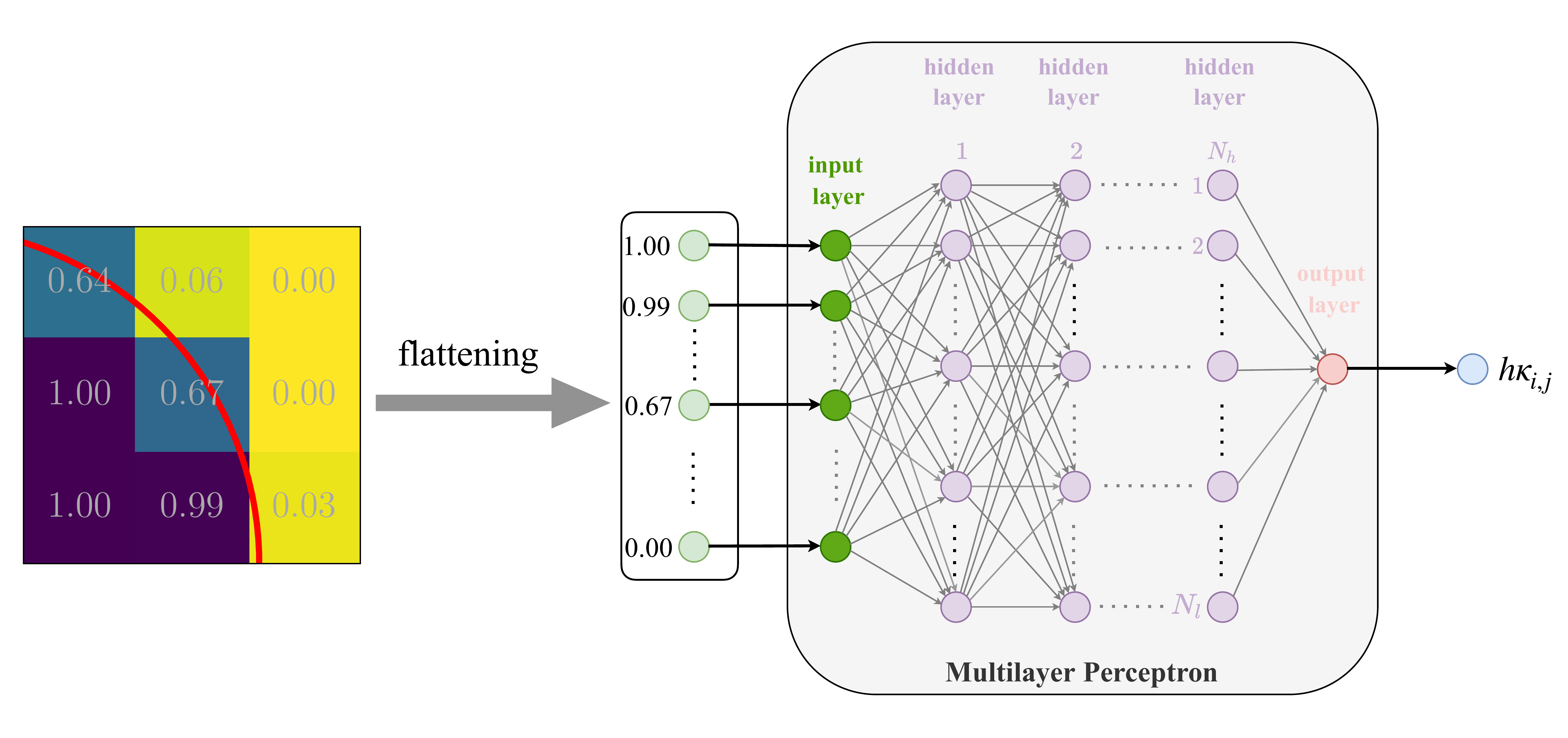}
\end{center}
\caption{\label{fig:MLPStd} Standard multilayer perceptron (StdMLP) model to estimate the curvature of the interfacial segment in the cell (i,j).}
\end{figure}

{\color{black}We consider a two-dimensional Cartesian domain partitioned with uniform square cells of side $h$}. The objective is to estimate the non-dimensional curvature $h\kappa_{i,j}$ from volume fractions data  in a local $3\times3$ block around the cell ($i,j$), i.e., {\color{black}  $ \vec A_{i,j}:=\{\alpha_{i+m,j+n}, \forall m,n=\{-1,0,1\}\}$.} This defines a nonlinear mapping:
 \begin{equation}
 \label{eq:mlpMap}
 	h \kappa_{i,j}=f_{\kappa}(\mathbf A_{i,j})=f_{\kappa}\begin{pmatrix}
 		\alpha_{i-1,j-1},&\alpha_{i,j-1},&\alpha_{i+1,j-1},\\
 		\alpha_{i-1,j},&\alpha_{i,j},	&\alpha_{i+1,j},\\
 			\alpha_{i-1,j+1},&\alpha_{i,j+1},&\alpha_{i+1,j+1}\end{pmatrix}.
 \end{equation}
An MLP model is developed to build this mapping. The model receives flattened volume fractions data and returns the non-dimensional curvature, cf.  figure~\ref{fig:MLPStd}. Small $3\times3$  stencil has been the standard choice for machine learning models \cite{ Qi2019-mu, Liu2021-sh, Larios-Cardenas2021-yf,Larios-Cardenas2021-sh, Buhendwa2021-qd}. The stencil configuration determines the dimension of the input layer, i.e., $N_i= 3\times 3=9$.  Next, the dimension of each hidden layer  along with the total number of hidden layers  and their activations functions have to be decided. These are the hyperparameters of MLP with great flexibility to configure. {\color{black} The model without any symmetry considerations will be referred to as the standard MLP (StdMLP) model hereafter.} 

\subsection{Symmetry considerations}\label{sec:symMLP}

 The curvature of an interface is invariant under symmetry transformations of input volume-fractions grid such as reflection over $x$ axis, reflection over $y$ axis, reflection over the diagonal lines at $y=x$ and $y=-x$.  Additionally, the absolute value of the predicted curvature, $|\kappa_{i,j}|$, should be independent of sign conventions. Namely, the curvature should only change its sign when the two fluids are swapped, not its magnitude. This is essential to maintain the geometry of anti-symmetric interfaces. An example is a sine wave where the curvature is maximum at its peak ($\kappa_{max}$), and minimum at its trough ($-\kappa_{\max}$). A good model should deliver the same magnitude, i.e., $|\kappa_{i,j}|=\kappa_{max}$, at these two extremal locations.
 
 \begin{figure}[t]
\begin{center}
\includegraphics[scale=0.31]{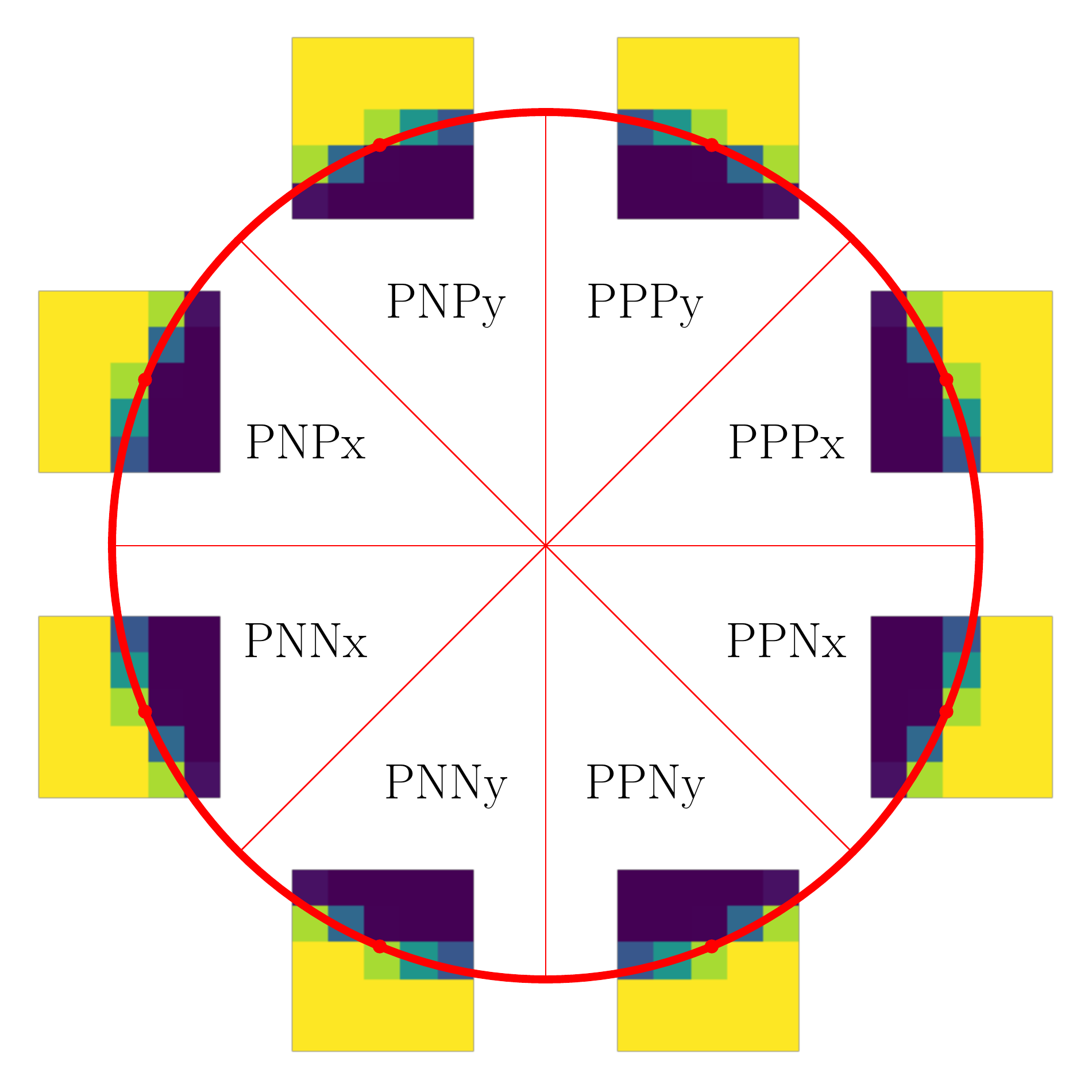}
~~~~~~~~~~~~~\includegraphics[scale=0.30]{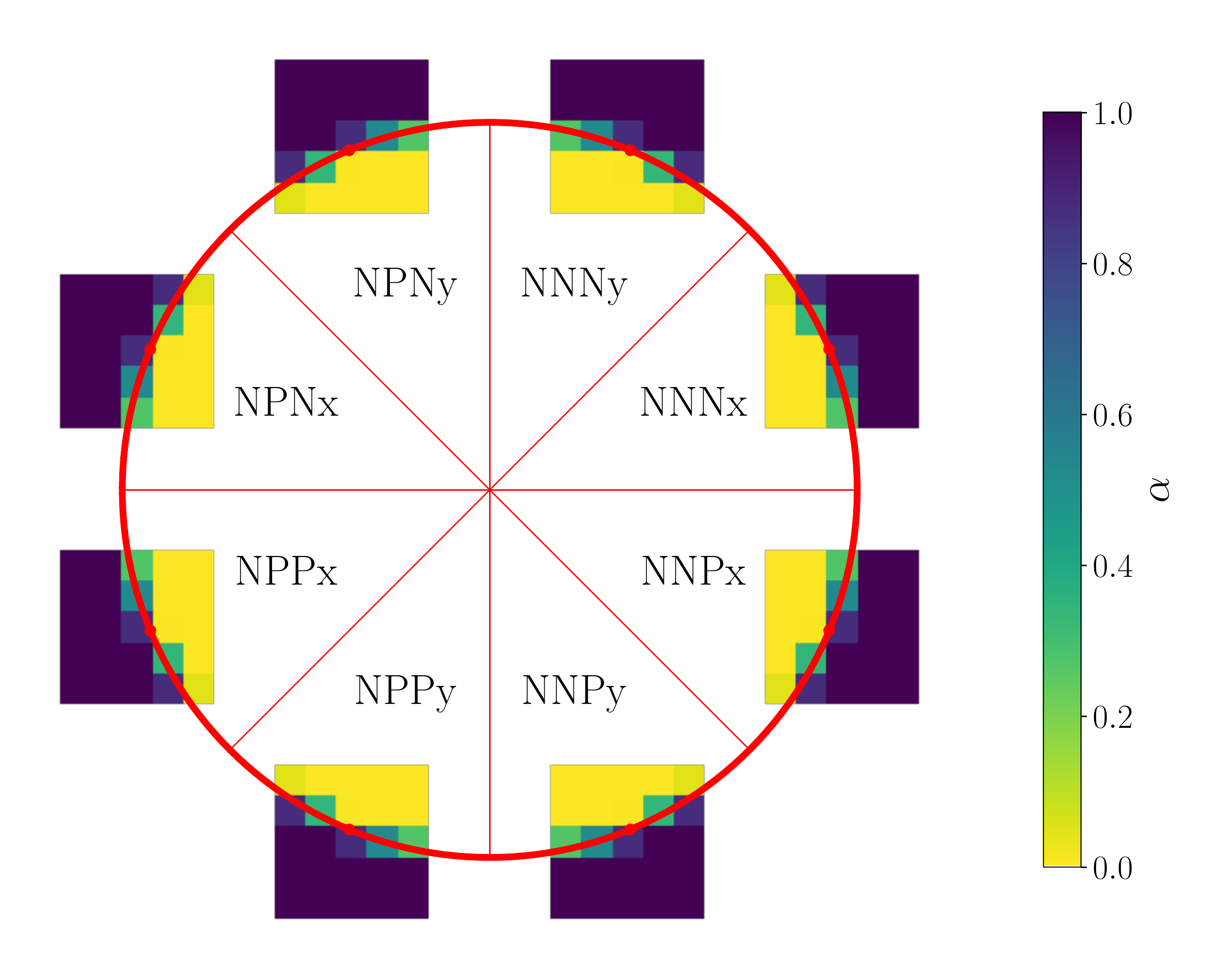}
\end{center}
\caption{\label{fig:2T} Geometrical zones of a circular interface, which are determined by the sign of the curvature and normal vector. See the text for the naming convention. }
\end{figure}

 In total, there are 16 symmetry configurations, or zones, in the plane. Figure~\ref{fig:2T} illustrates them for a circular interface.  The naming convention is in order: sign of the curvature ($\sgn \kappa$), sign of the $x$-normal ($\sgn n^x$), sign of the $y$-normal ($\sgn n^y$), the direction of the maximal normal $n_{i,j}^{\max}:=\max\{|n_{i,j}^x|,|n_{i,j}^y|\}$. For instance, PPPx represents the interface zone with $\kappa>0$, $n^x>0$, $n^y>0$ and $n_{i,j}^{\max}=|n_{i,j}^x|$. Let us define the following transformations on  volume fraction data:
 \begin{enumerate}
	\item Vertical reflection over $x$-axis: 
	\begin{equation}\label{eq:gx} g_x:\vec A_{i,j}\mapsto \begin{pmatrix}
 		\alpha_{i-1,j+1},&\alpha_{i,j+1},&\alpha_{i+1,j+1},\\
 		\alpha_{i-1,j},&\alpha_{i,j},	&\alpha_{i+1,j},\\
 			\alpha_{i-1,j-1},&\alpha_{i,j-1},&\alpha_{i+1,j-1}\end{pmatrix},\end{equation} 
	\item Horizontal reflection over $y$-axis:
	\begin{equation}\label{eq:gy} g_y:\mathbf A_{i,j}\mapsto \begin{pmatrix}
 		\alpha_{i+1,j+1},&\alpha_{i,j+1},&\alpha_{i-1,j+1},\\
 		\alpha_{i+1,j},&\alpha_{i,j},	&\alpha_{i-1,j},\\
 			\alpha_{i+1,j-1},&\alpha_{i,j-1},&\alpha_{i-1,j-1}\end{pmatrix},\end{equation} 
	\item Diagonal reflection over the line that is at $\theta=\pi/4$ angle from $x$-axis:  
	\begin{equation}\label{eq:grot} g_{rot}:\mathbf A_{i,j}\mapsto \begin{pmatrix}
 		\alpha_{j-1,i-1},&\alpha_{j,i-1},&\alpha_{j+1,i-1},\\
 		\alpha_{j-1,i},&\alpha_{j,i},	&\alpha_{j+1,i},\\
 			\alpha_{j-1,i+1},&\alpha_{j,i+1},&\alpha_{j+1,i+1}\end{pmatrix},\end{equation}
	\item Swapping the fluids:   
	 \begin{equation} \label{eq:gswap} g_{swap}:{\mathbf A}_{i,j} \mapsto \vec J-\mathbf A_{i,j},\end{equation}
\end{enumerate}
where $J$ represents an array with all elements being unity.  Given an  $\mathbf A_{i,j}$ in one of the zones, corresponding symmetries in other zones can be generated using the transformations  Eqs. (\ref{eq:gx}--\ref{eq:gswap}). A symmetry-preserving curvature estimator then should deliver the same curvature magnitude in all these zones, e.g., $f_\kappa(\mathbf A_{i,j})=f_\kappa(g_{rot}(\mathbf A_{i,j}))=f_\kappa(g_{x}(\mathbf A_{i,j}))=f_\kappa(g_{y}(\mathbf A_{i,j}))$ and $f_\kappa(\mathbf A_{i,j})=-f_\kappa(g_{swap}(\mathbf A_{i,j}))$.

{\color{black} The training of the symmetry-preserving MLP model, denoted as SymMLP}, are restricted to the interfacial segments where the curvature is positive and the maximal normal is in $y$ direction, i.e., PPPy, PNPy, PPNy and PNNy.  If the input data is in one of the remaining symmetry zones with positive curvature then it is transformed to its counterpart in training zones using $g_{rot}$ operator Eq.~(\ref{eq:grot}). In a simulation environment, this requires the information of normal vector $n_{i,j}^{\max}$ on runtime. In VOF-PLIC method, the normal vector is explicitly available from the interface reconstruction step. This couples the accuracy of curvature estimation to the accuracy of interface reconstruction scheme. This drawback of our approach is shared by the height function method, which uses  $n_{i,j}^{\max}$  to orient the height function. 

Next, we need a strategy to cover the negative-curvature zones. To enforce magnitude-invariance with respect to fluid swapping, we propose the following steps. First,  the volume fractions are normalized:
\begin{equation}
\vec{\widetilde A}_{i,j} = 2\mathbf A_{i,j}-\vec J
\label{eq:normal}
\end{equation}
The normalization modifies the range of input data from $[0,1]$ to $[-1,1]$, and fluid-swapping operation becomes
	 \begin{equation} \label{eq:gswap2} \widetilde g_{swap}:\vec {\widetilde{\vec A}}_{i,j} \mapsto -\vec {\widetilde{\vec A}}_{i,j}. \end{equation}
 Using the normalization, magnitude-invariance with respect to fluid swapping can now be ensured if the MLP mapping $f_\kappa$ has odd symmetry, i.e., $f_\kappa(\widetilde {\vec A}_{i,j})=-f_\kappa(\tilde g_{swap}(\widetilde{\vec A}_{i,j}))=-f_\kappa(-\widetilde{\vec A}_{i,j})$. An MLP with this property can be developed using: 
\begin{enumerate}
\item Bias-free neurons, e.g., for a single artificial neuron or perceptron (\ref{eq:neuron}): $\hat y =\phi(\mathbf w \cdot \hat {\mathbf x})$,
\item Activation functions with odd symmetry such as hyperbolic tangent function, i.e., $\phi=\tanh$.
\end{enumerate}
It is easy to show the odd symmetry in an artificial neuron: $\hat y =\tanh(\mathbf w \cdot \hat {\mathbf x})=-\tanh(-\mathbf w \cdot \hat {\mathbf x})$. Thanks to the composition structure of MLP (\ref{eq:MLP}), the odd symmetry in individual neurons extends to the whole network and $f_\kappa$ becomes:
\begin{equation}
\begin{aligned}
h\kappa_{i,j}&=W_{o}^T(\tanh(W_{N_h}^T(\ldots (\tanh(W_{1}^T\widetilde{\vec{A}}_{i,j})\ldots)),\\
				&=-W_{o}^T(\tanh(W_{N_h}^T(\ldots (\tanh(-W_{1}^T\widetilde{\vec{A}}_{i,j})\ldots )),
\end{aligned}
\end{equation}
where we have employed linear activation function at the output layer. Because of odd-symmetry, an MLP model trained in positive-curvature zones has identical performance in negative curvature-zones.  Removing bias parameters from neurons potentially reduces the learning capacity of MLPs. Nevertheless, we observed good results with odd-symmetric MLPs. 

\begin{figure}[!t]
\begin{center}
\includegraphics[scale=0.165]{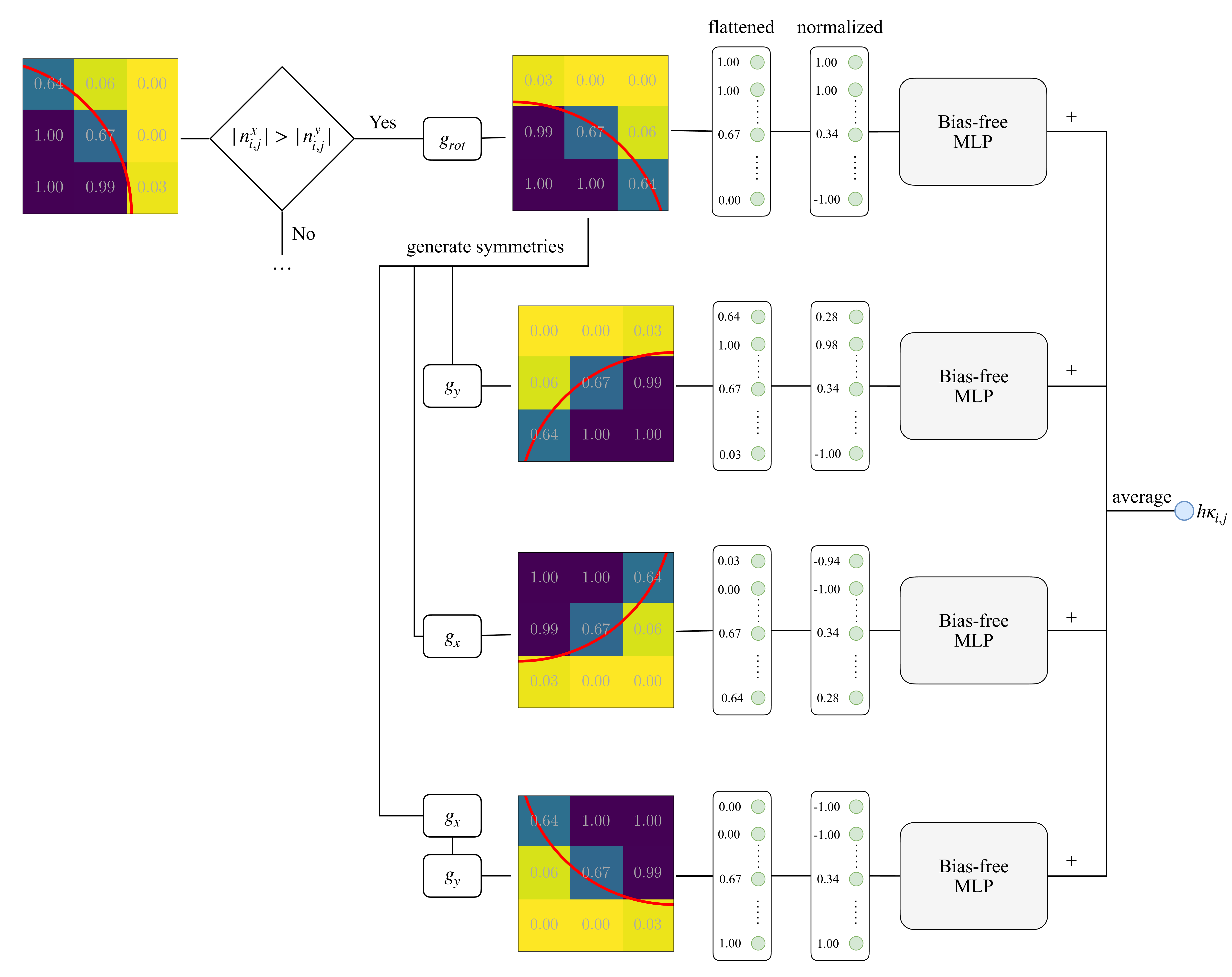}
\end{center}
\caption{\label{fig:MLPSym} Symmetry-preserving multilayer perceptron (SymMLP) model to estimate the curvature of the interfacial segment in the cell ($i,j$).}
\end{figure}

In the final step of our approach, the outcome of four symmetry zones are averaged. The overall structure of the symmetry-preserving MLP model is illustrated in figure~\ref{fig:MLPSym}. 

\section{Model building}\label{sec:model}

\subsection{Generation of synthetic datasets}\label{sec:datasets}
The training datasets are generated from circular interface segments. The curvature of a circular interface with a radius $R$ is given by $\kappa=\pm1/R$ with the sign depending on the arrangement of volume fractions.  We observed best results when $R$ is drawn from a log-uniform distribution, i.e.: 
\begin{equation}
	\text{$R=2^p$, ~~~~with~~~ $\mathcal P(p)=\frac{1}{p^{\max}-p^{\min}}$},
\end{equation}
where $\mathcal P$ is the probability distribution function, $p^{\min} $ and $p^{\max}$ are the lower and upper limits of the random exponent $p$. We choose $p^{\min} =1$ and $p^{\max}=12$. The second random parameter to generate a circular arc is the angle $\theta$ for which we employ a uniform distribution in $\theta\in [0,2\pi)$. 

\begin{figure}[!t]
\begin{center}
\includegraphics[scale=0.22]{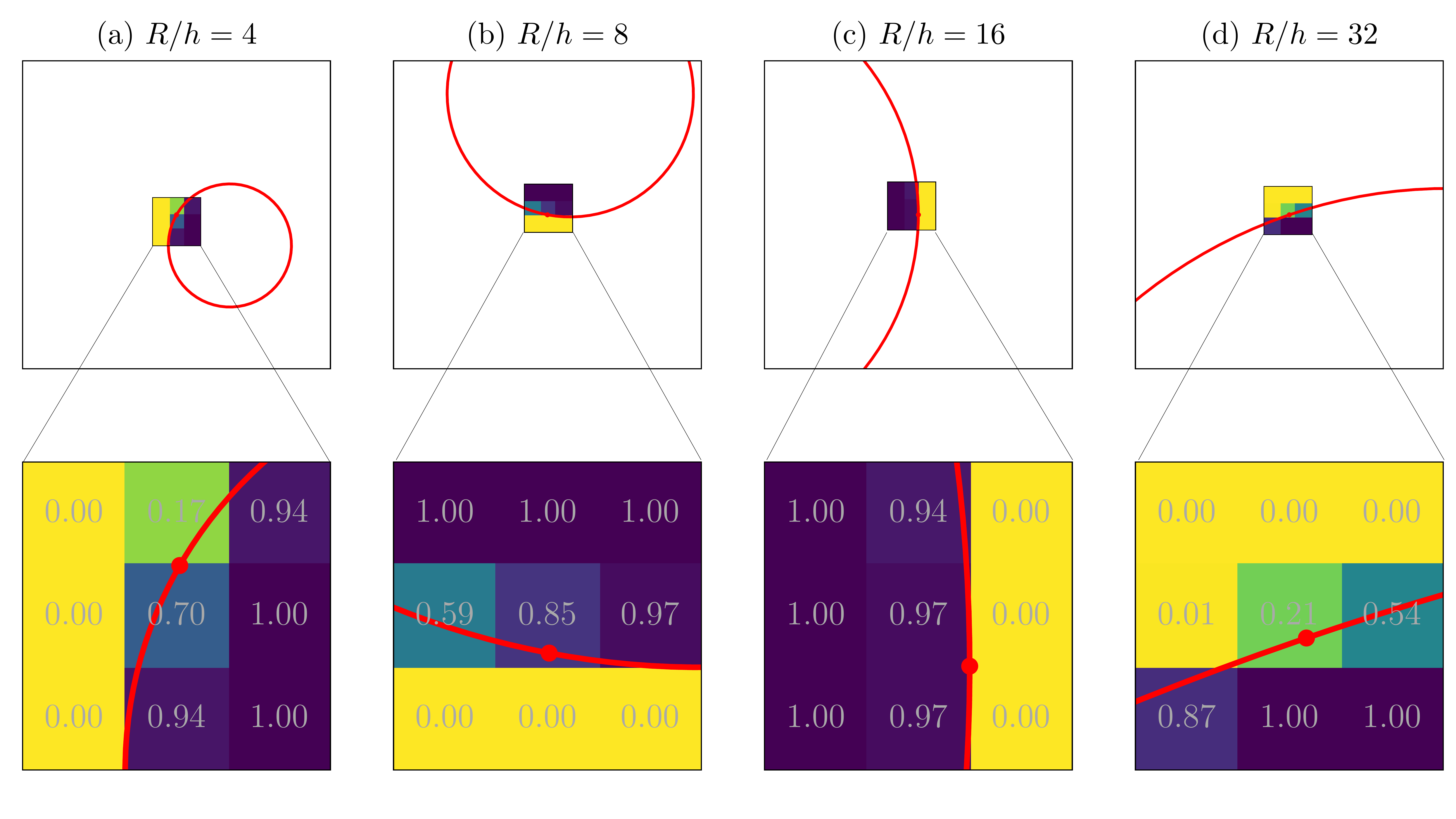}
\end{center}
\caption{\label{fig:samples} Samples from training dataset with varying curvature using circles of different size, i.e., $R/h=\{4,8,16,32\}$.}
\end{figure}

 {\color{black}We consider a uniform structured grid with spacing $h=1$.} An individual sample of the dataset can be generated using the following routine:
\begin{enumerate}
	\item Draw a radius $R$ from log-uniform distribution.
	\item Draw an angle $\theta$ from uniform distribution.
	\item Identify the centroid $\mathbf x_{i,j}$ of the cut cell hosting the point $\vec x(R,\theta)$, and generate $3\times 3$ grid of cells around this cut cell. 
	\item Obtain the volume fractions $\vec A_{i,j}$ by partitioning each cell into $200\times200$ subcells and numerically integrating the phase indicator function $\chi$ with Simpson's rule. Here, $\chi=1$ and $\chi=0$ inside and outside the circle, respectively. 
	\item If negative-curvature samples are considered, randomly swap fluids with $g_{swap}(\vec A_{i,j})$ to allow samples with curvature $\kappa_{i,j}=-1/R$.
	\item If zone restriction is considered, calculate the normal vector $\mathbf n(\theta)$ and flip indices with $g_{rot}(\vec A_{i,j})$ when  $n^x>n^y$.
	\item If normalization is considered, normalize input volume fractions by Eq.~\ref{eq:normal}.
	\item Add the input $\vec A_{i,j}$, or $\widetilde{\vec A}_{i,j}$, and output $\kappa_{i,j}$ to the dataset.
\end{enumerate}
 Several illustrative samples are demonstrated in figure~\ref{fig:samples}.  The dataset for the standard MLP model StdMLP is generated through the steps~1-5, and 8. The symmetry-preserving MLP model SymMLP is built from the dataset generated via steps~1-4,~6-8. Fairly large datasets with $5\times 10^5$ samples are generated.  Each dataset is partitioned into two parts with ratios $0.7$ and $0.3$ for training and validation purposes, respectively.

\subsection{Training of models}
\begin{figure}[!t]
\begin{center}
\includegraphics{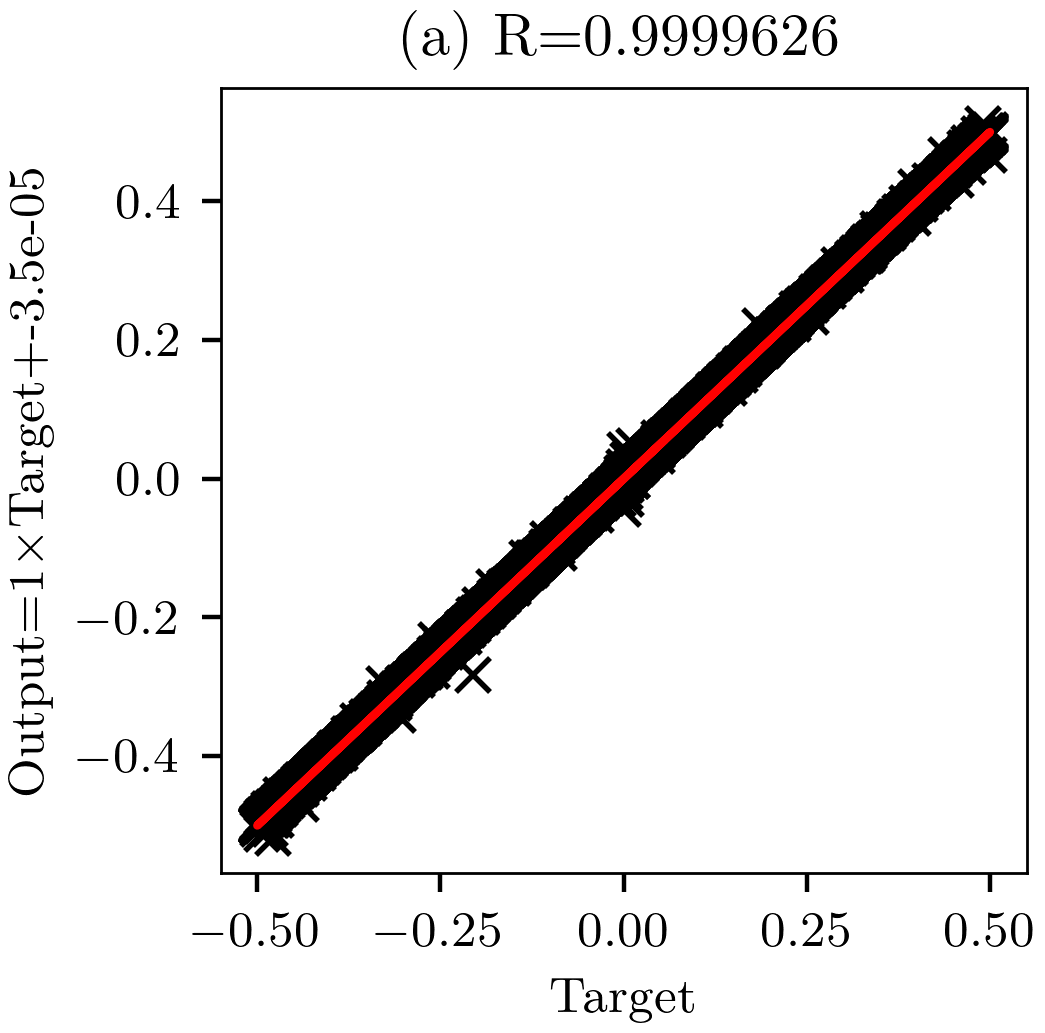}
~~~~~~~~~~~~\includegraphics{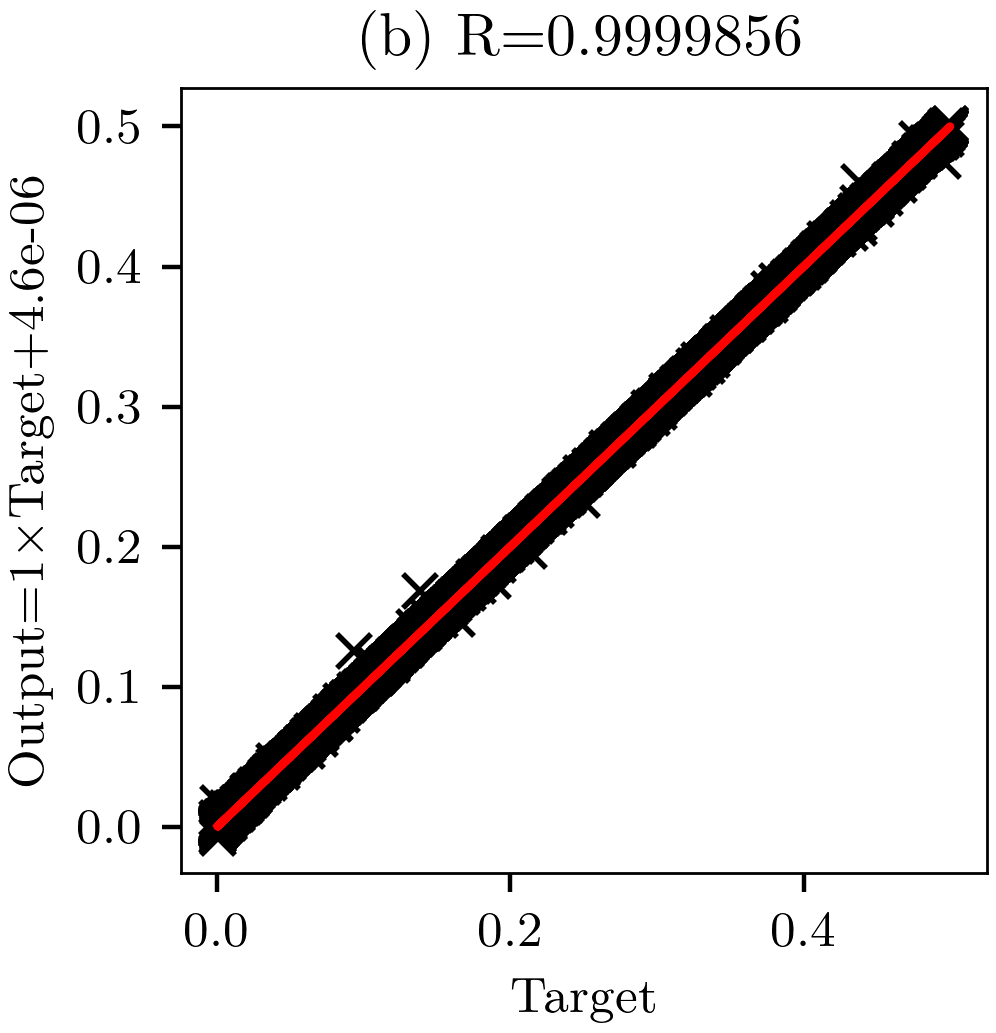}
\end{center}
\caption{\label{fig:KrefTrain} Performance of MLP models on their datasets. (a) StdMLP. (b) SymMLP. Red line represents the linear regression fit. The correlation coefficient $R$ is listed at the top of the plots. }
\end{figure}

Once the datasets are available,  hyperparameters of the neural network, i.e., activation functions, architecture, optimizer, learning rate, batch size and epocs, have to be tuned.  For activation functions, $\tanh$ function is employed in hidden layers to preserve the odd symmetry as discussed in Sec~\ref{sec:symMLP}. The output layer has the linear activation function.  Next, a suitable architecture have to be found for hidden layers. Several deep architectures varying from $8\times 4$ (4 layers with 8 neurons) to $120\times4$ are tested. $30\times 4$ provided good balance between accuracy and cost. Weights of the neurons are obtained by minimizing the mean-square error between the predictions and reference values. This is done using TensorFlow v2.5.0. A minimum is sought  with Adam optimization algorithm  \cite{Kingma2014-kj}.  The learning rate is initially set to $10^{-3}$ and gradually reduced with an exponential decay rate of 0.99. We employ a mini-batch optimization approach with a batch size of 512. The training is run for 2000 epocs. During the training, the optimizer randomly selects the batches in each epoch. Because of this inherent stochasticity,  a different local minimum may be obtained in each training attempt despite the identical input and network configuration.  Consequently, some scatter is observed in the performance. For each model configuration, we have repeated the whole training process about 100 times and selected best performing StdMLP and SymMLP models. The performance of the models over their datasets are presented in figure~\ref{fig:KrefTrain}. The curvature extends to negative values in StdMLP (figure~\ref{fig:KrefTrain}a) as this model include all 16 symmetry zones. SymMLP delivers a better model fit thanks to its dataset focusing only on 4 symmetry zones (figure~\ref{fig:KrefTrain}b). It should be noted that averaging over 4 symmetry zones is not yet applied at this training stage.

\section{A-priori analysis: analytical tests}\label{sec:testing}

\begin{figure}[!t]
\begin{center}
\includegraphics{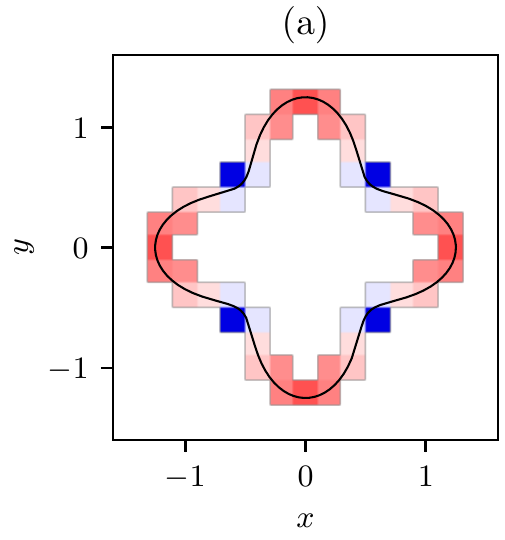}
~~~~~~~~\includegraphics{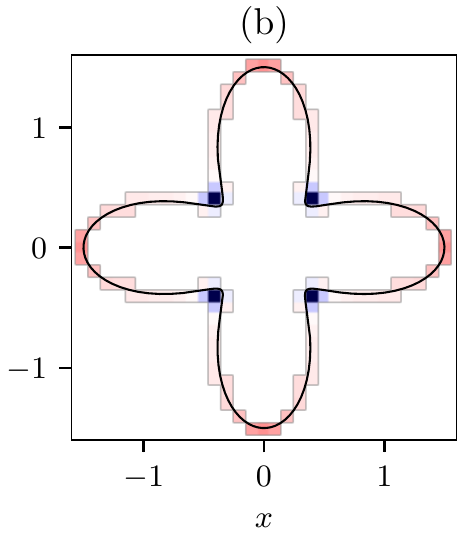} 
\includegraphics{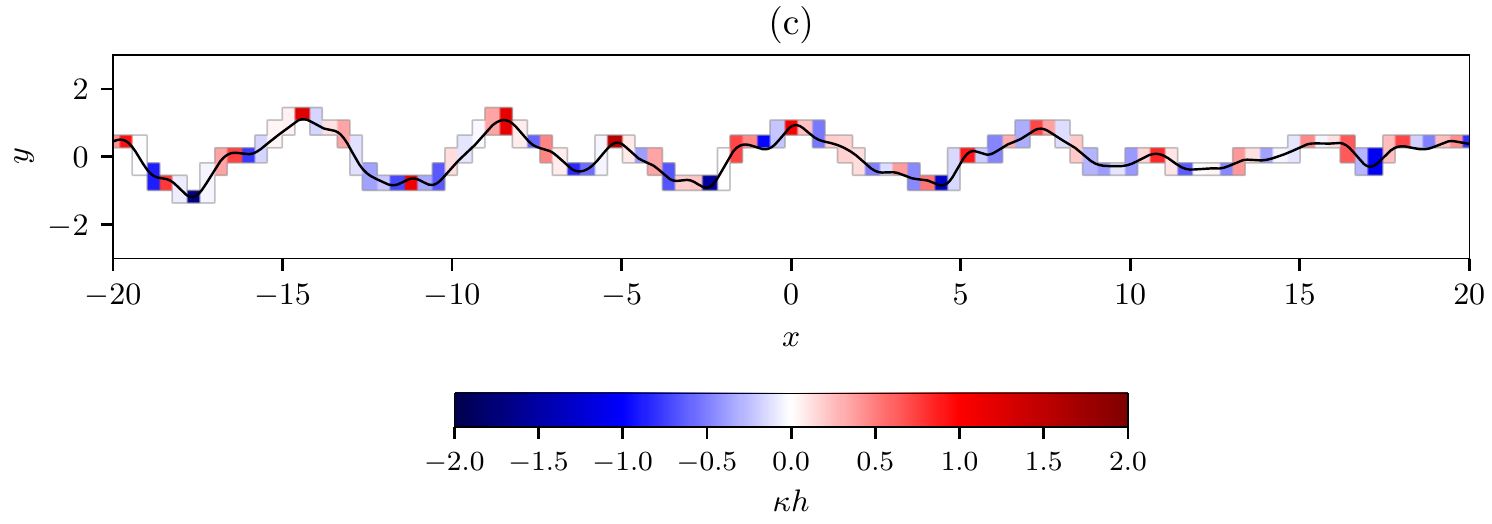}
\end{center}
\caption{\label{fig:tests} Test cases for curvature estimation. Contours show the non-dimensional curvature in cells. (a) 4-petalled star with $R_s=1$, $A_s=0.25$ and $n=4$. Resolution of contours is $R_s/h=5$.  (b) 4-petalled star with $R_s=1$, $A_s=0.5$ and $n=4$ with resolution $R_s/h=10$. (c) Random wave with Pierson--Moskowitz spectrum with significant wave height $H_s=1.5$ m and peak-energy wavelength $\lambda_p=10$ m for resolution $\lambda_p/h=100$. }
\end{figure}

The developed models are first tested on shapes where the curvature is analytically available. The considered test cases are shown in figure~\ref{fig:tests}. The first two test cases are based on the star geometry defined as:
\begin{equation}
	\mathbf x(s) = \begin{bmatrix}  (R_s+A_s\cos(n_s s))\cos s \\ (R_s+A_s\cos(n_ss))\sin s  \end{bmatrix}, ~\forall s\in[0,2\pi],
\end{equation} 
Two 4-petalled stars ($n_s=4$ and $R_s=1$) with one smooth ($A_s=0.25$) and one more acute ($A_s=0.5$) interface are selected, cf. figures~\ref{fig:tests}a,b. The final test case is irregular wavy interface representing 2D ocean waves found on fully-developed seas. To this end, we consider a random wave field that is composed of multiple Fourier components
\begin{equation}
	y(x)=\sum A_n\cos(k_nx+\phi_n), 
	\label{eq:Hwave}
\end{equation}
where the phases $\phi_n\in [0,2\pi]$ are randomly selected from uniform distribution and the amplitudes are specified by
\begin{equation}
\frac{1}{2}\left | A_{n}\right |^2=\Psi(k_n)\Delta k_n=\sqrt{\frac{g}{k_n}}G(\omega)\Delta k_n.
\label{eq:An}
\end{equation}
In this expression, $G(\omega)$ is the Pierson--Moskowitz wave spectrum~\cite{pierson1964proposed}:
\begin{equation}
G(\omega)= \gamma H_s^2 \frac{\omega_p^{4}}{\omega^{5}}\exp\left[-\frac{5}{4} \left(\frac{\omega}{\omega_p}\right)^{-4}\right]
\label{eq:G}
\end{equation}
where $H_s$ is the significant wave height, $\omega_p$ is the peak angular frequency, and the constant $\gamma$ is used to adjust the energy of the spectrum to significant wave height. The random wave field can be determined by two parameters: $k_p$ and $H_s$. For convenience, we will use peak wavelength $\lambda_p=2\pi/k_p$ instead of wavenumber. We set $\lambda_p=10$ m and $H_s=1.5$ m. The selected realization for testing is depicted in figure~\ref{fig:tests}c. {\color{black}In total, 200 wave components  with wavelengths $2\pi/k_n\in\{0.1 \lambda_p,0.2 \lambda_p,\ldots,20 \lambda_p\}$ are used to generate the waveform using Eq.~(\ref{eq:Hwave}). The amplitudes of the components are obtained from Eq.~(\ref{eq:An}) where we set $\Delta k_n=40\pi/\lambda_p$. In Eq.~(\ref{eq:G}), the conversion to spatial variables is done using the dispersion relation for deep water waves: $\omega$=$\sqrt{gk_n}$.}

The standard MLP model of Qi et al.~\cite{Qi2019-mu}, tagged as QLSZT2019 hereafter, is also included in benchmarks. These authors employed a single hidden layer with 100 neurons.  From conventional schemes, we tested the height function method, which employs a large $3\times 7$ stencil, and will be denoted as HF37. Height functions are aligned with the maximal normal direction $n_{i,j}^{\max}$. In this regard, SymMLP model and HF37 method require the information of normal vector to detect  $n_{i,j}^{\max}$. Poor estimation of this quantity might yield additional errors in both methods. To account for these uncertainties, we will employ Youngs algorithm \cite{youngs1982time} to estimate the normal vector instead of using analytically available one. Youngs algorithm is a simple and fast method, which is commonly employed as precursor to more advanced interface reconstruction methods \cite{Aulisa2007}. In this method, the normal is calculated using the discrete gradient ($\nabla_h$) of the volume fraction field: $\vec n_{i,j}=-\nabla_{h} \alpha_{i,j}$, i.e.,  
\begin{align}
\label{eq:nx}
 	n^x_{i,j}&=\frac{1}{8h}\left[\alpha_{i+1,j+1}-\alpha_{i-1,j+1}+2\alpha_{i+1,j}-2\alpha_{i-1,j}+\alpha_{i+1,j-1}-\alpha_{i-1,j-1}\right], \\
 \label{eq:ny}
 	n^y_{i,j}&=\frac{1}{8h}\left[\alpha_{i+1,j+1}-\alpha_{i+1,j-1}+2\alpha_{i,j+1}-2\alpha_{i,j-1}+\alpha_{i-1,j+1}-\alpha_{i-1,j-1}\right],
 \end{align}
where $h$ is the uniform grid spacing. 

The error of the curvature models is evaluated by $L_2$ norm defined as
 \begin{equation}
 \label{eq:L2}
 	E_2=\sqrt{\frac{\sum^{N_i}(\kappa-\kappa_{ref})^2}{N_i}},
 \end{equation}
 and by $L_\infty$ norm given by
  \begin{equation}
   \label{eq:LInf}
 	E_\infty= \max\{\left|\kappa-\kappa_{ref}\right|\} ,
 \end{equation}
 where $\kappa$ is the estimated curvature in a cut cell, $\kappa_{ref}$ is the reference curvature, and $N_i$ is the number of interface cells where $\varepsilon<\alpha_{i,j}<1-\varepsilon$.   We set the threshold to $\varepsilon=10^{-3}$. 

\begin{figure}[!t]
\begin{center}
\includegraphics{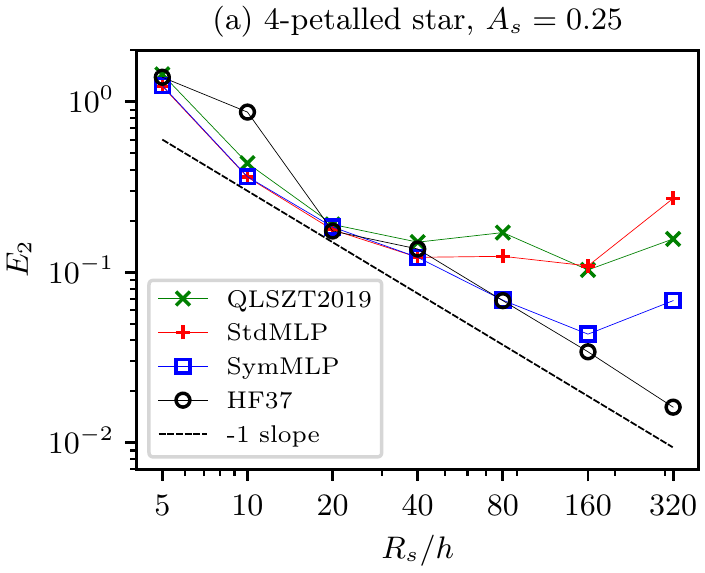}
~~~~~~\includegraphics{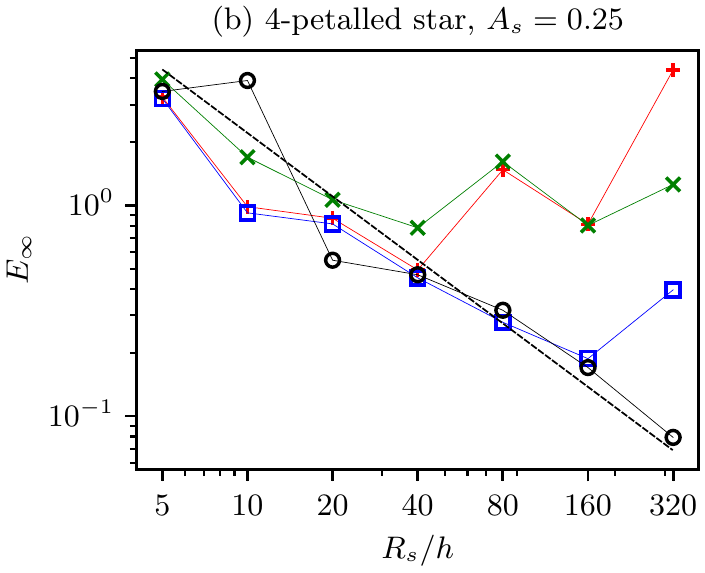}
\includegraphics{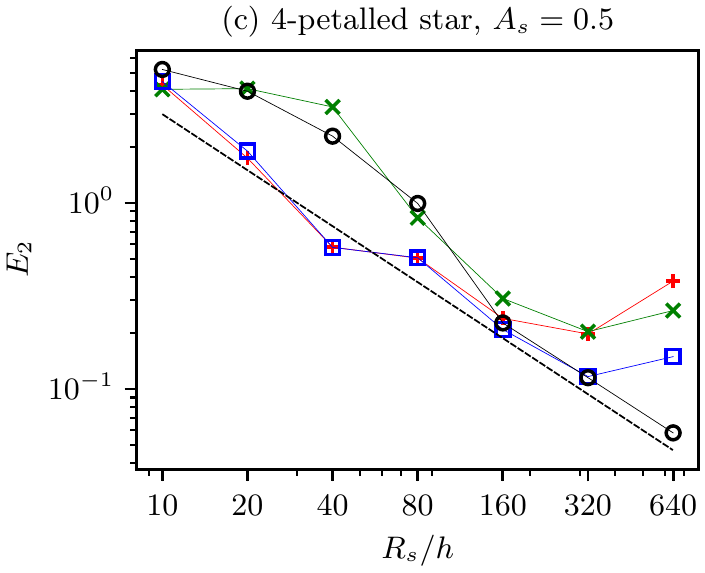}
~~~~~~~\includegraphics{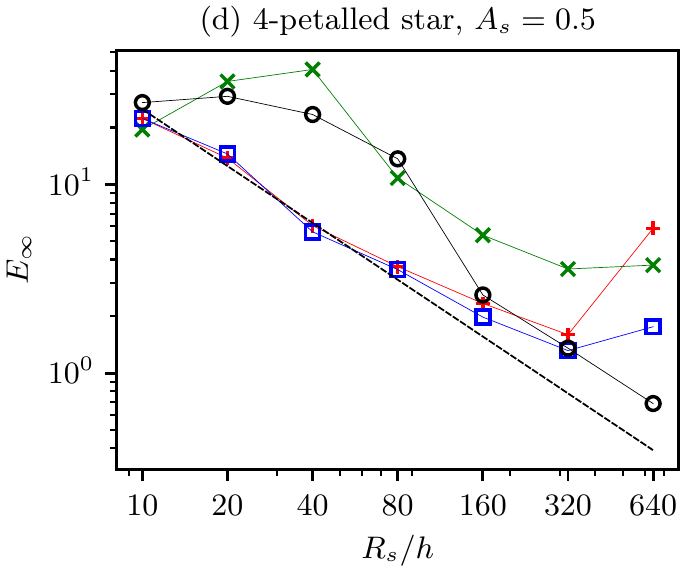}
\includegraphics{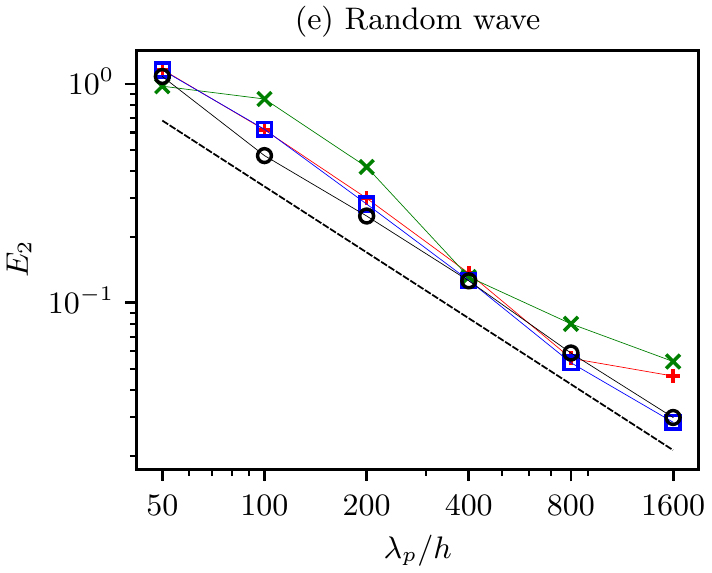}
~~~~~~~\includegraphics{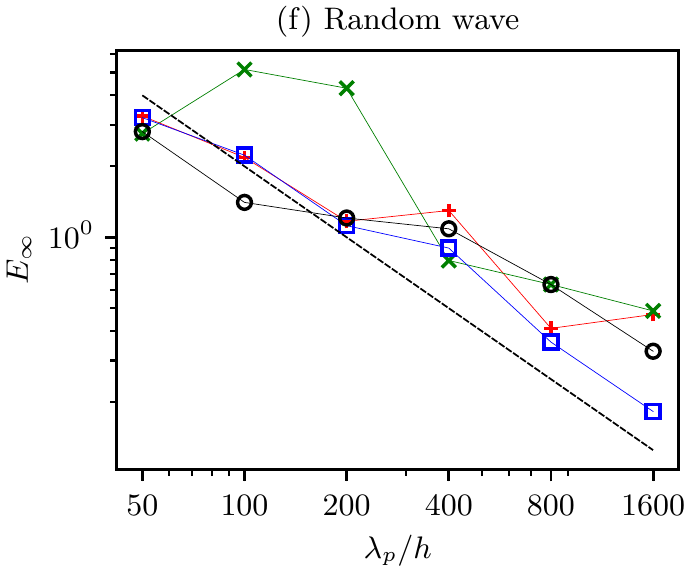}
\end{center}
\caption{\label{fig:convStar} Convergence of MLP models with grid refinement for star geometries with $R_s=1$ and $n_s=4$. Estimation error is evaluated by  (a,c,e): $L_2$ norm (\ref{eq:L2}); (b,d,f): $L_\infty$ norm (\ref{eq:LInf}).}
\end{figure}

Figure~\ref{fig:convStar} compares the convergence of models. We observe that standard MLP benefits from a deeper $30\times 4$ architecture. Except on finest grids, StdMLP have somewhat superior accuracy compared to QLSZT2019 model with $100\times 1$ architecture.  While SymMLP and its standard counterpart StdMLP behave similarly on coarse resolutions, SymMLP have significantly more accurate estimations when the grid is refined. Thus, the new symmetry-preserving approach allows a more robust MLP model that delivers reliable estimations on a wider range of resolutions. The accuracy of HF37 remains below SymMLP on coarse to moderate resolutions, in particular for the acute star, cf. $R_s/h=10$ to $160$ in figures~\ref{fig:convStar}c,d.

\section{A-posteriori analysis: simulation performance}\label{sec:simulations}    
MLP models can be implemented into CFD codes to estimate the curvature of arbitrarily shaped interfaces in simulations. Open-source CFD code OpenFOAM v2006 \cite{ofv2006} is selected for this purpose.  For benchmarking, the height function method is implemented into OpenFOAM library as well. We further include the RDF method by Scheufler et al. \cite{scheufler2021twophaseflow,gamet2020validation} in our benchmarks.  Finally, the curvature estimation with the stock two-phase flow solver of OpenFOAM, \textit{interFoam}, is also considered. \textit{interFoam} employs a variant of algebraic VOF method to transport the interface. The OpenFOAM implementations are elaborated in Appendix~A.

 At the beginning of each time step, we first identify the cells whose volume fraction satisfy $\delta <\alpha_{i,j}<1-\delta$ with $\delta \sim 10^{-6}-10^{-8}$ as cut cells. The interface is then reconstructed in these cut cells. To this end, first, the interfacial normal is estimated with Mixed--Youngs-Centered (MYC) scheme by Aulisa et al.~\cite{Aulisa2007}. After finding its orientation, the interfacial segment in each cut cell is positioned to represent the correct volume fraction value using the iterative algorithm of Scheufler and Roenby \cite{Scheufler2019}.  Finally, once the velocity field is obtained, the interface is advected with \textit{isoAdvector} algorithm \cite{Roenby2016}. 
 
 All MLP models provide poor estimations when a cut cell predominantly occupied by one fluid.  In this configuration, the cutplane is furthest from the cell center where the curvature is evaluated. {\color{black} We have observed improvements when MLP estimation is switched off in such cells, and curvature is set to zero, i.e., $\kappa_{i,j}=0$}. Therefore, henceforth,  the estimation of curvature will be restricted to cells with  $\varepsilon<\alpha_{i,j}<1-\varepsilon$ where $\varepsilon>\delta$ when MLP models are considered. The results are sensitive to $\varepsilon$, and we obtained the best results with $\varepsilon=10^{-3}$. 

\subsection{Stationary bubble and spurious currents}\label{sec:staticBubble}

Let us consider a static cylindrical bubble with a circular interface of diameter $D$ separating two fluids with equal density ($\rho$) and kinematic viscosity ($\nu$). The momentum equation for this stationary problem reduces to:
\begin{equation}
\label{eq:Laplace}
0=-\nabla P+\sigma \kappa \nabla \chi,
\end{equation}
where $P$ is the pressure field. This equation dictates an equilibrium between surface tension forces and pressure forces. Satisfying this balance numerically has been a long-lasting challenge \cite{popinet2018numerical}. 
Inaccurate curvature gradients act like a source term driving artificial production of vorticity \cite{Abadie2015-rd}, and spurious currents are generated in this process. Previous works have shown that these capillary artefacts are generated in the beginning of the simulation. They eventually decay and reach an equilibrium state in which the intensity of currents depends on the estimated curvature fields \cite{Popinet2009}. 

Two characteristic timescales are significant: (i) capillary timescale: $t_{\sigma}=\sqrt{\rho D^3/\sigma}$; (ii) viscous timescale: $t_{\nu}=D^2/\nu$. A characteristic non-dimensional number, Laplace number, is defined using the ratio of these timescales:
\begin{equation}
La=\left (\frac{t_\nu}{t_\sigma}\right)^2=\frac{\sigma D}{\rho \nu^2}.
\end{equation}
The viscous velocity scale of the problem is $U_\mu=\sigma/\mu$. The intensity of spurious currents can be quantified by the maximum capillary number 
\begin{equation}
Ca_{\max}(t):=\frac{U_{\max}(t)}{U_{\mu}}=\frac{\mu U_{\max}(t)}{\sigma},
\label{eq:CaMax}
\end{equation}
where $U_{\max}$ is the maximum velocity in the domain. 

\begin{figure}[!t]
\begin{center}
\subfloat[]{\includegraphics[scale=0.19]{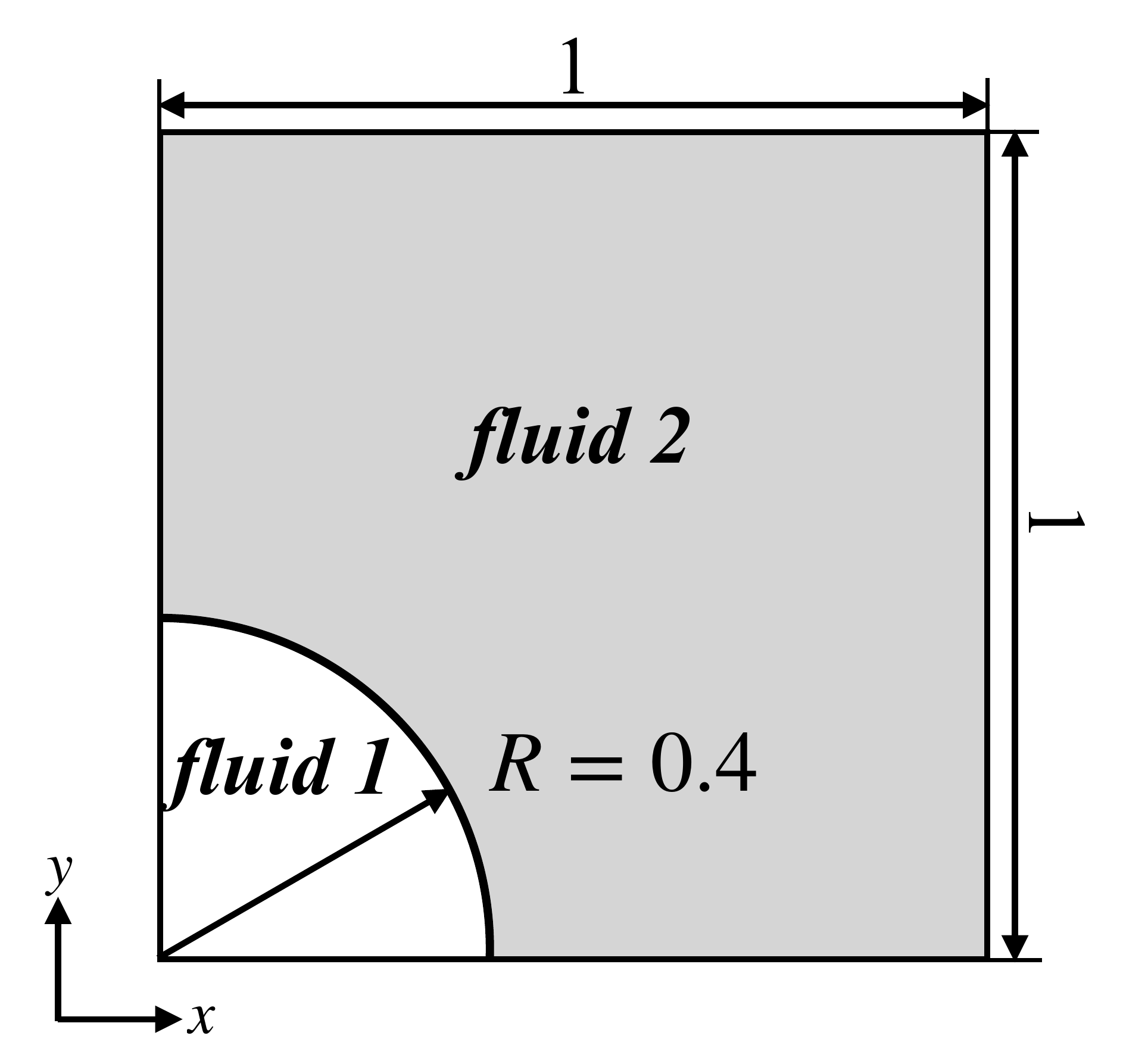}}
\subfloat[]{ ~~~~~~\includegraphics[scale=0.18]{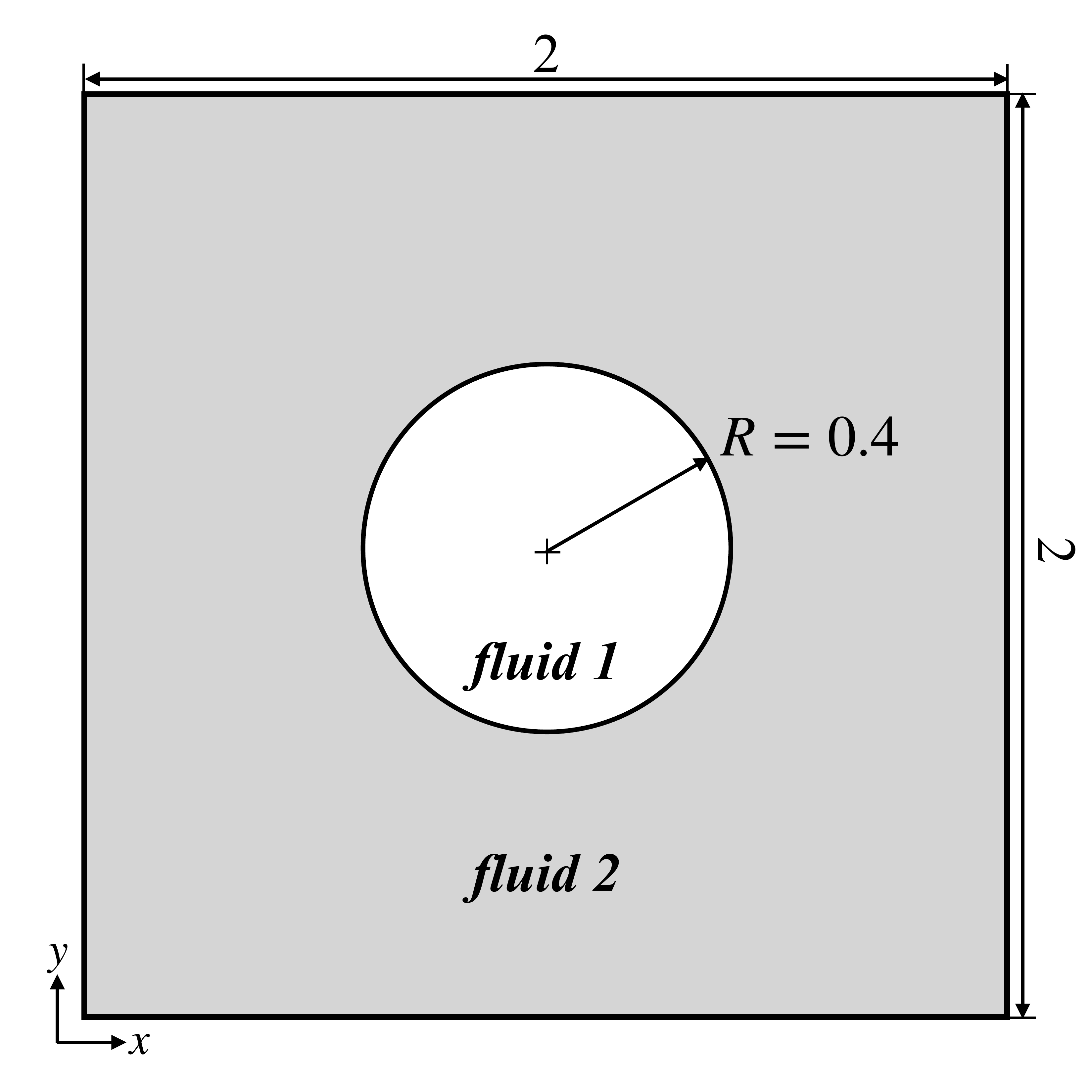}}
\caption{\label{fig:staticBubble} Flow configurations for the static bubble tests. (a) Quadrant. (b) Full circle. Symmetry boundary conditions are imposed at all boundaries.}
\end{center}
\end{figure}

We consider two different computational configurations, cf.~figure~\ref{fig:staticBubble}. The first set-up is a standard configuration in literature \cite{gamet2020validation,Abadie2015-rd,Popinet2009}, where only one quadrant of the circle is considered, and symmetry conditions are imposed on the boundaries. The second configuration considers the whole circle. The domains are partitioned into uniform square cells with a resolution of $N\times N$ and $2N\times 2N$. We observed that changes in the Laplace number only modify the decay horizon of the spurious currents, not their intensity in the equilibrium range. Therefore, all tests are done with a fixed Laplace number of $La=12000$. A time-stepping approach blending  backwards Euler and Crank--Nicolson schemes is used for time integration. Hereby, the weights of the Euler and Crank-Nicolson methods are set to 0.1 and 0.9, respectively.

\begin{figure}[!t]
\begin{center}
\includegraphics{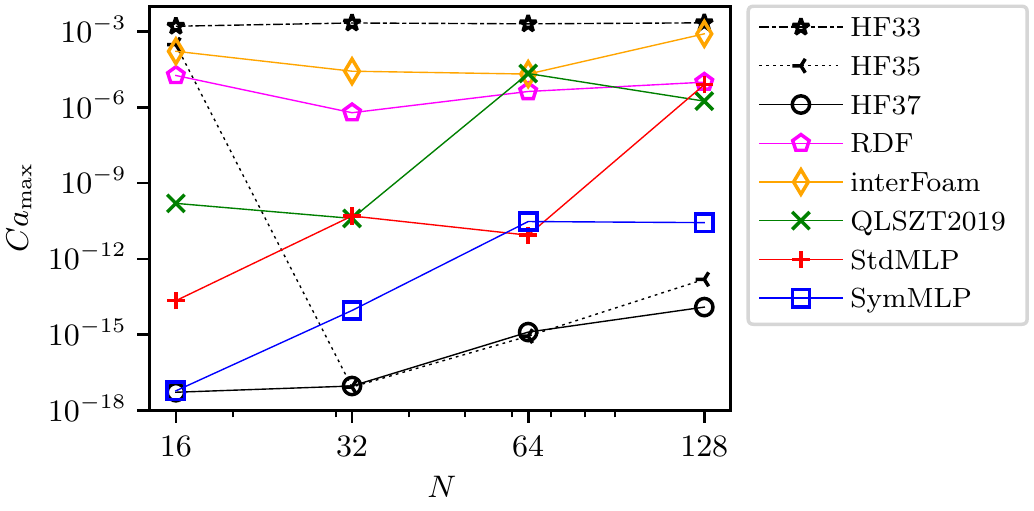}
\caption{\label{fig:CaMaxConv} Convergence of the maximum amplitude of the spurious currents with grid refinement in the quadrant domain.}
\end{center}
\end{figure}

Figure~\ref{fig:CaMaxConv} presents the equilibrium values of $Ca_{\max}$ in the quadrant domain for different grid resolutions. The results for the height function method using smaller stencils (HF33 and HF35 using $3\times 3$ and $3\times 5$ stencils respectively) are also presented. The simulations with HF37 are completely free from spurious currents on all grid resolutions within the limits of machine precision. This implies that a perfect discrete balance between the pressure gradient and surface-tension forces is obtained.  The ability of HF37 to remove spurious currents was previously reported by other investigators, e.g., Ref.~\cite{Popinet2009}. This key property is reproduced here, which validates our implementation of the height function method. This ability of the height function method is completely lost when a small $3\times 3$ stencil is used. HF33 generates the most intense spurious currents among all methods. Extending the stencil to $3\times 5$ dramatically improves the performance of the height function methods but HF35 still struggles on coarse $N=16$ resolution. The best performing MLP model is SymMLP which yields low spurious currents on all tested grids with values below $Ca_{\max}< 10^{-10}$. Among standard MLP models, StdMLP estimations generate less intensive spurious currents. Both  StdMLP and QLSZT2019 models perform poorly on the finest  $N=128$ resolution yielding results in the same range with the RDF method.  

\begin{figure}[!t]
\begin{center}
\includegraphics{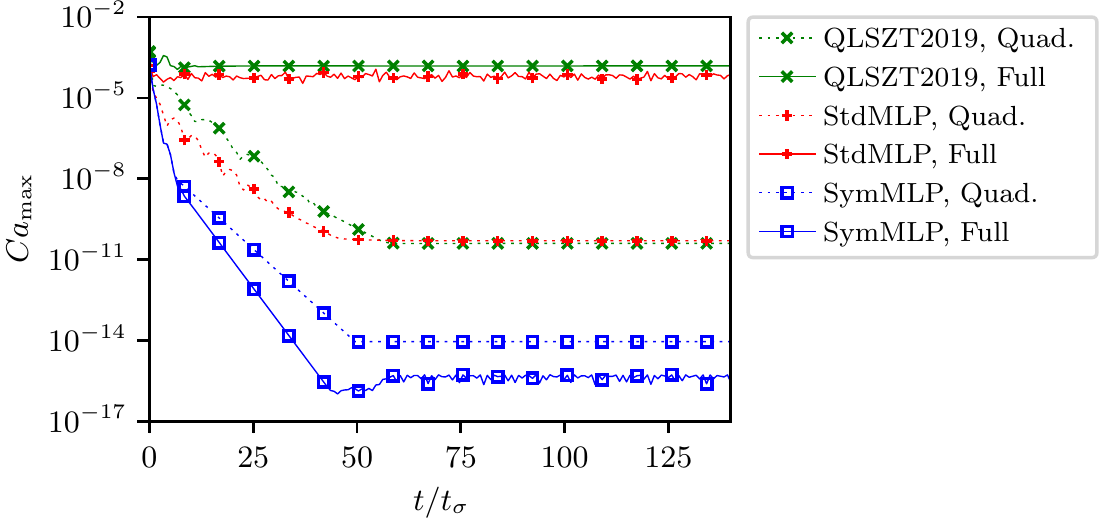}
\caption{\label{fig:CaMaxFull} Temporal evolution of the maximum amplitude of the spurious currents in the full-circle and quadrant domains is shown for MLP models on a resolution $N=32$.}
\end{center}
\end{figure}

\begin{figure}[!t]
\begin{center}
\includegraphics{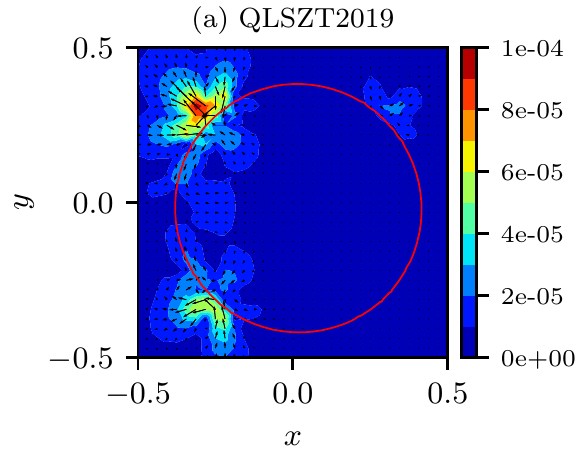}
\includegraphics{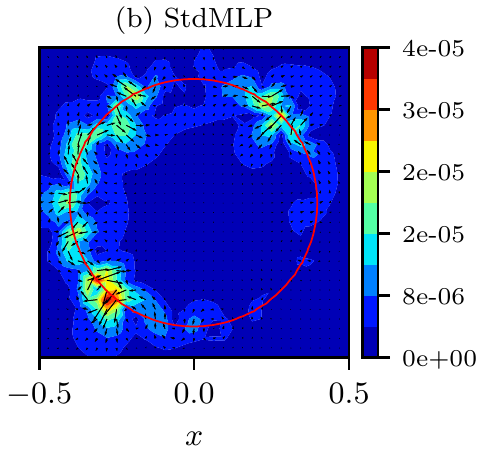}
\includegraphics{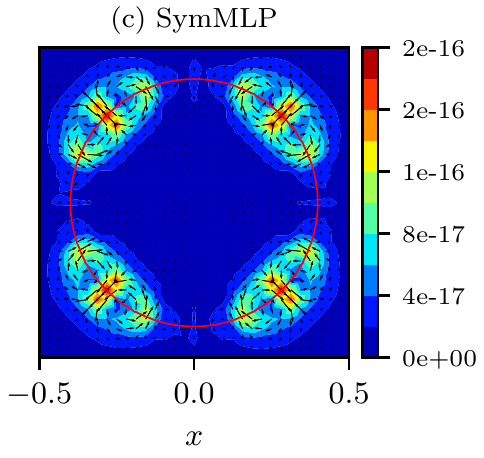}
\caption{\label{fig:uFull} 
Spurious currents in full-circle domain tests on $N=32$ resolution. Filled contours present the magnitude of the velocity $|\mathbf u|$ and arrows demonstrate its direction.  }
\end{center}
\end{figure}

 The quadrant domain explicitly enforces reflectional symmetries through symmetry boundary conditions on the left and bottom boundaries. Therefore, it does not test how well these symmetries are preserved in the remaining quadrants.  The domain containing the whole circle (figure~\ref{fig:staticBubble}b) has to be utilized to fully examine the models. Figure~\ref{fig:CaMaxFull} compares $Ca_{\max}(t)$ in the full-circle and quadrant domains for MLP models on $N=32$ resolution. Interestingly, the performance of symmetry-preserving model improves in the full-circle domain where the steady-state value of $Ca_{\max}$ is a few decades lower than its counterpart in the quadrant domain. In contrast, standard MLP models greatly struggles in the big domain, and spurious currents settle into equilibrium at significantly higher values. Velocity fields in the equilibrium range are plotted in figure~\ref{fig:uFull} for the full-circle domain. It is clear that standard MLP models fail to preserve symmetries whereas SymMLP exhibit quasi-identical quadrants, hence perfectly capturing all reflectional and diagonal symmetries.

\subsection{Standing capillary wave}
The second test case with dynamic interface is a small-amplitude standing wave. An initial sinusoidal perturbation $A_0\cos(kx)$ superimposed on a flat interface located at $y=0$ between two fluids at rest. The oscillations develop under the effect of surface tension and are gradually damped by viscosity. This is a standard benchmark case, e.g. Refs.~\cite{Popinet2009,Owkes2015-tn, scheufler2021twophaseflow,gamet2020validation}, with a reference solution for vanishing amplitudes provided by Prosperetti \cite{Prosperetti1981-fd}.

Temporal dimension is scaled by the normal mode frequency $\omega_0$, which is defined by the dispersion relation $\omega_0^2=\sigma k^3/2\rho$.  Both fluids have equal density  ($\rho$) and kinematic viscosity ($\nu$). The Laplace number is $\mathrm{La}:=\sigma \lambda/\rho\nu^2=3000$ where $\lambda=2\pi/k$ is the wavelength of the perturbation. 
Following Ref. \cite{Popinet2009}, we employ a computational domain with a size $[-\lambda/2,\lambda/2]\times[-3\lambda/2,3\lambda/2]$. The domain is periodic in horizontal direction and slip boundary conditions are applied at top and bottom boundaries. Implicit Euler method is used for time stepping. Time step size is constrained by a small Courant number $Co=5\times10^{-4}$ to prevent capillary instabilities. 

 \begin{figure}[!t]
\begin{center}
\includegraphics{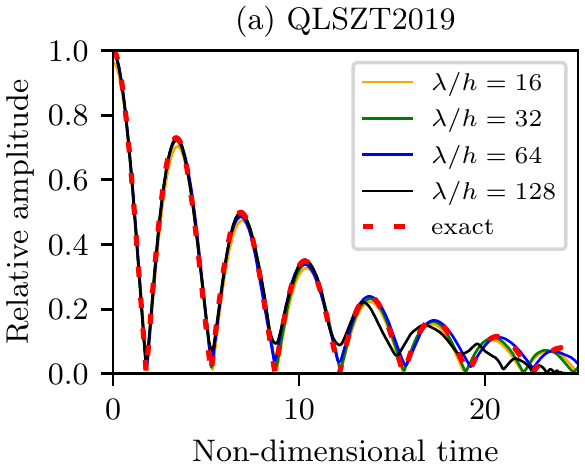}
\includegraphics{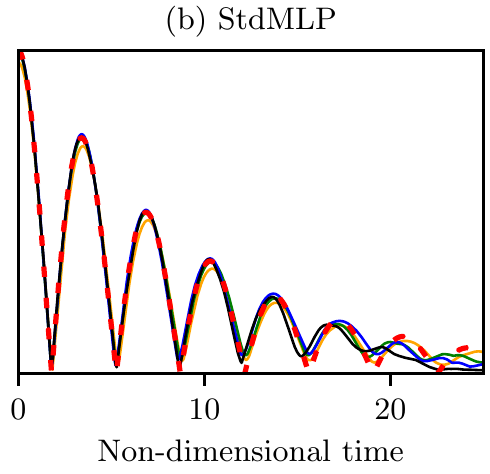}
\includegraphics{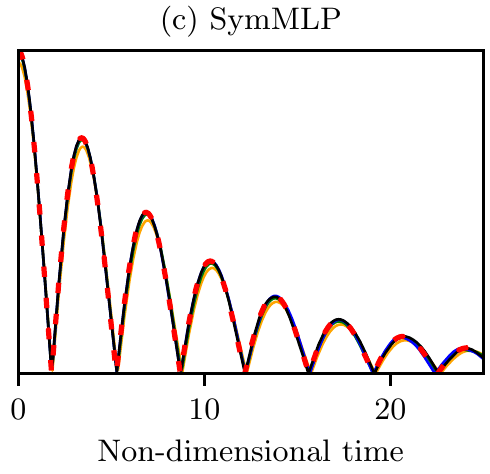}

\end{center}
\caption{\label{fig:scwTime} Time histories of maximum interface elevation $\eta^{\max}/A_0$ of the capillary wave simulated with different MLP curvature estimation models on different resolutions.}
\end{figure}

\begin{figure}[!t]
\begin{center}
\includegraphics[]{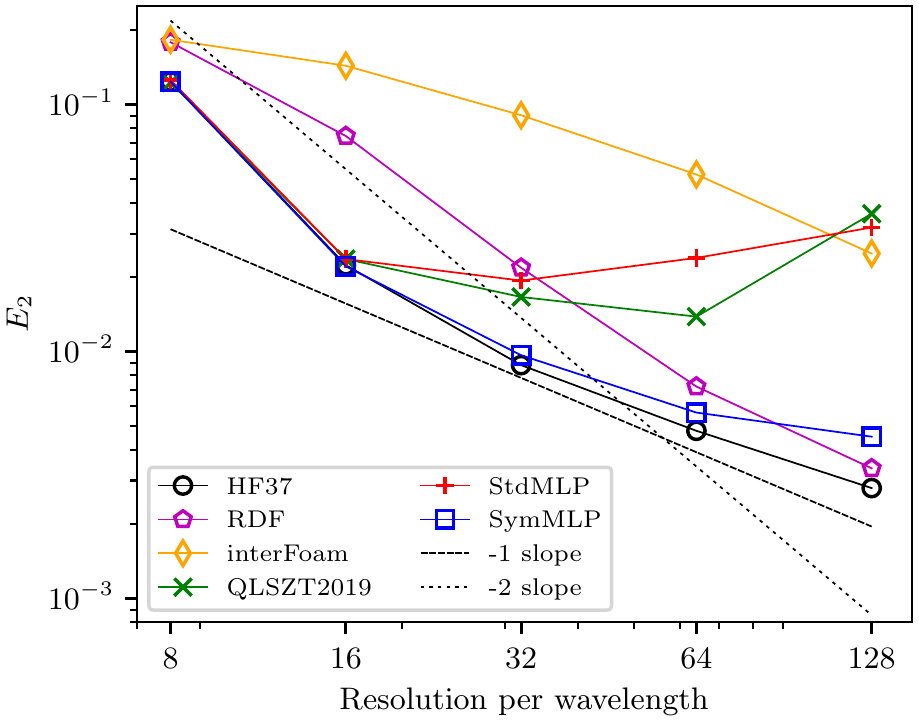}
\end{center}
\caption{\label{fig:cw1} Convergence results for the capillary wave on $L_2$ norm  (\ref{eq:Eeta}). (a): $\rho_1/\rho_2=1$, $\nu_1/\nu_2=1$ and $La=3000$.}
\end{figure}

Figure~\ref{fig:scwTime} shows the time histories of maximum interface elevations $\eta^{\max}(t)=\max\{|\eta(x,t)|\}$ for MLP models on different resolutions. These results are compered to the analytical solution of Prosperetti. Standard schemes perform similarly and struggle in reproducing the wave characteristics after $t>10$. Gradual deteriorations result in remarkable discrepancies towards the end of the event. This is not the case for symmetry-preserving model SymMLP. It exhibits a vastly superior accuracy and reproduces the capillary oscillations on all presented grids. 

We can quantify the errors in time histories with the $L_2$ norm 
\begin{equation}
    E_2(t;\eta)=\frac{1}{A_0}\sqrt{\frac{\omega_0}{25}\int_0^{25/\omega_0}(\eta^{\max}(t)-\eta^{\max}_{ref}(t))^2\mathrm d t}
    \label{eq:Eeta}
\end{equation}
where $\eta^{\max}_{ref}$ are the exact maximum interface elevations. 
Figure~\ref{fig:cw1} presents the results. The convergence rate of interFoam is close to first order. RDF scheme appear to have a convergence rate between 1 and 2. HF37 initially converges in second order but gradually drops to first order and below. Convergence rates are much less applicable to MLP schemes. The convergence in standard schemes flattens quickly and eventually divergence in accuracy is observed. The symmetry-preserving model does not suffer divergence in the considered resolutions range.

\subsection{Rising bubble}
\begin{figure}[!t]
\begin{center}
\includegraphics[scale=0.33]{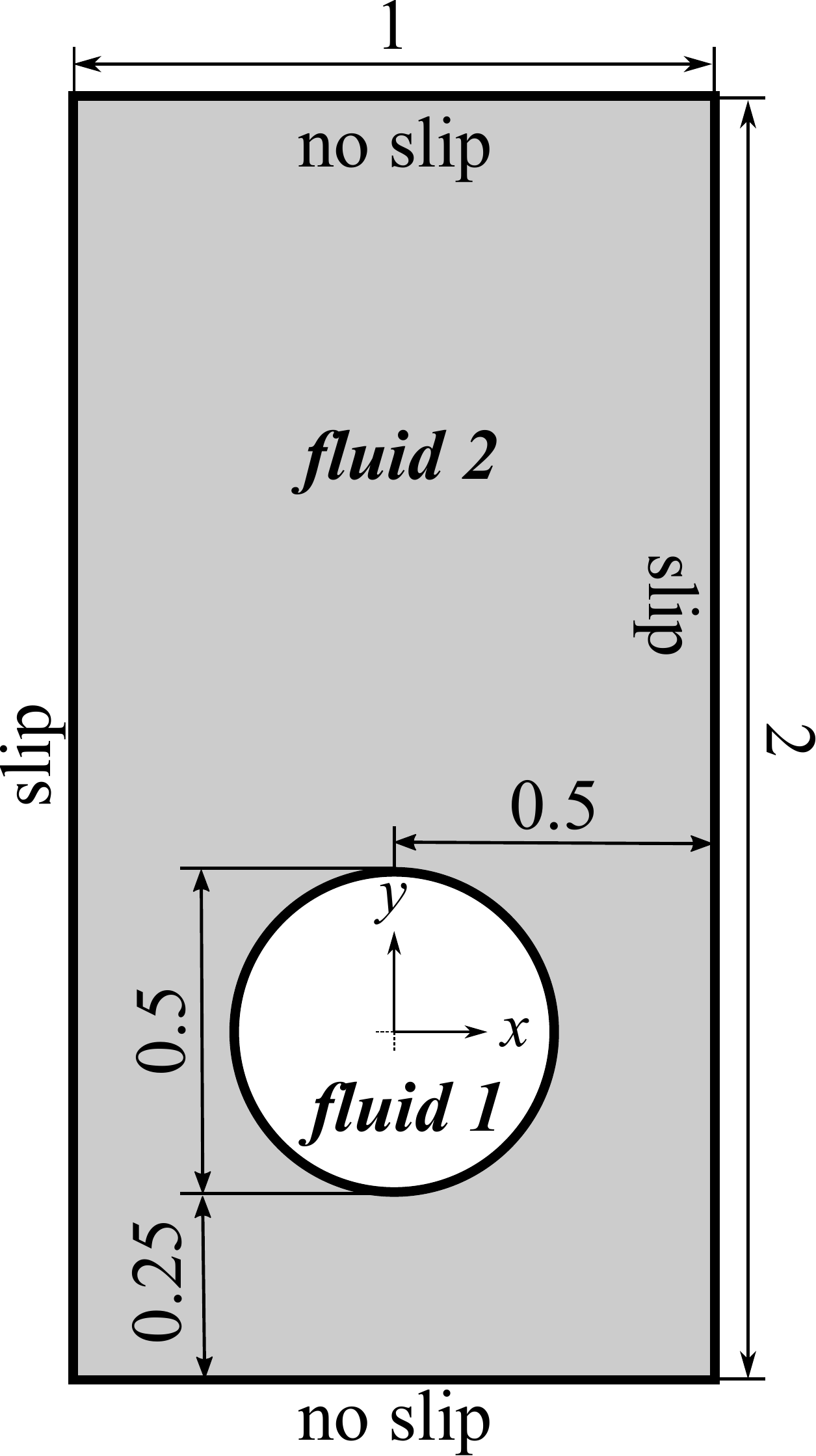}
\end{center}
\caption{\label{fig:hysing} Initial and boundary conditions for the rising bubble benchmark problem. Dimensions are given in centimetres. }
\end{figure}

The next benchmarking problem, designed by Hysing et al. \cite{Hysing2009-gn}, concerns with the rising of a two dimensional bubble in a liquid column. Figure~\ref{fig:hysing} depicts the problem configuration. The second test case in Ref.~\cite{Hysing2009-gn} is selected. This case is  characterized by high density and dynamic-viscosity ratios between fluids, and the bubble undergoes significant deformations while rising. The liquid phase has the physical parameters $\rho_1=1000$ kg m$^{-2}$, $\mu_1=10$ kg m$^{-1}$ s$^{-1}$, whereas the gas phase is defined with $\rho_2=1$ kg m$^{-2}$, $\mu_2=0.1$ kg m$^{-1}$ s$^{-1}$. The surface tension is $\sigma=1.96$ kg s$^{-2}$. The gravitational acceleration, pointing in $-y$ direction, is set to $g=0.98$ m s$^{-2}$. The dynamics of the flow is characterized by the Bond/E\"otv\"os number 
\begin{equation}
	\mathrm{Bo}=\frac{\rho_1 g D_0^2}{\sigma},
\end{equation}
and the Galilei number 
\begin{equation}
	\mathrm{Ga}=\frac{\rho_1 g^{1/2} D_0^{3/2}}{\mu_1},
\end{equation}
where $D_0$ is the diameter of the bubble in the initial configuration. We set $\mathrm{Bo}=125$ and $\mathrm{Ga}=35$, which puts the bubble in the peripheral break-up regime \cite{tripathi2015dynamics}.

A domain of $\Omega:=[-0.5,0.5] \times [-0.5,1.5]$ cm$^2$ is defined with uniform grid resolution $N \times 2N$, cf. figure~\ref{fig:hysing}. We set free-slip boundary condition at lateral boundaries and no-slip boundary condition at top and bottom boundaries. The maximum time step size is restricted by Courant number $\mathrm{Co}=0.025$. We vary the resolution $N$ between 80, 160, 320 and 640. A case with $N=2560$ resolution using HF37 will serve as a reference. We observe good grid convergence at this resolution. 

 \begin{figure}[!t]
\begin{center}
\includegraphics{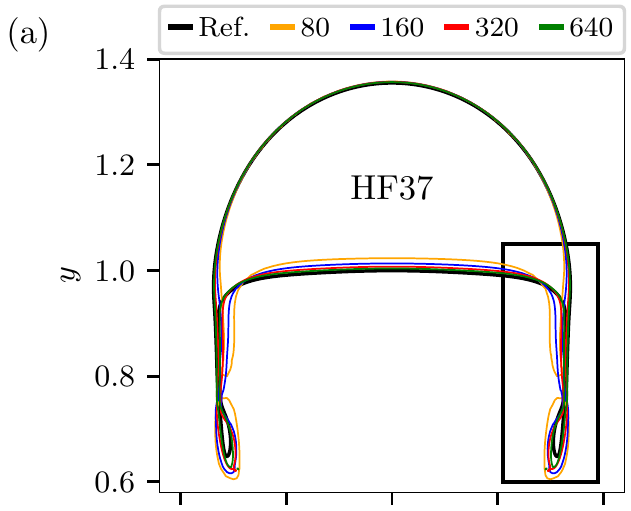}
\includegraphics{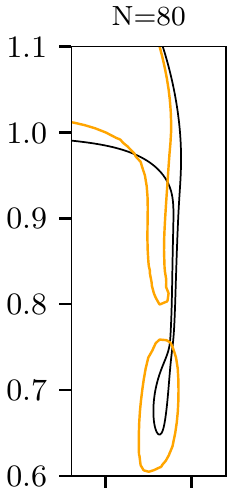}
\includegraphics{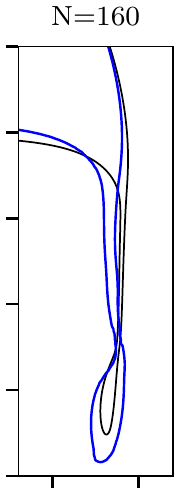}
\includegraphics{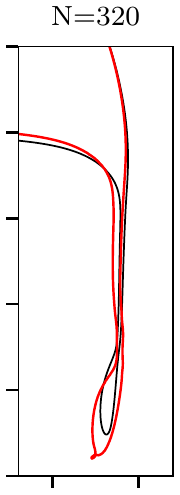}
\includegraphics{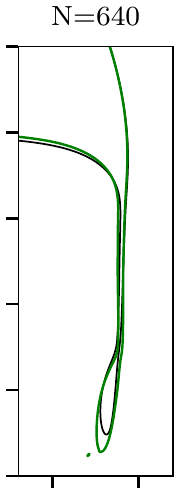}

\includegraphics{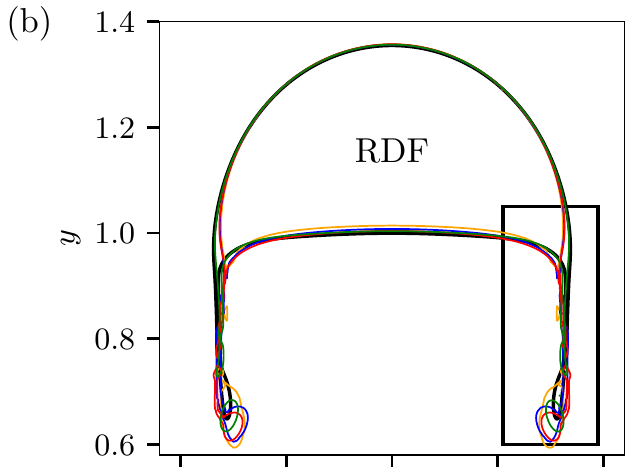}
\includegraphics{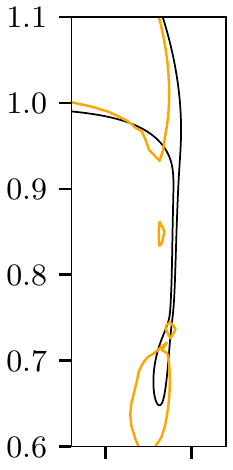}
\includegraphics{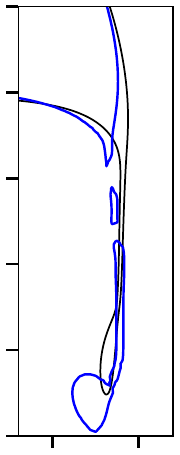}
\includegraphics{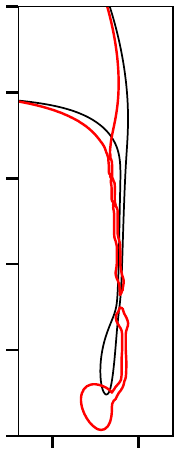}
\includegraphics{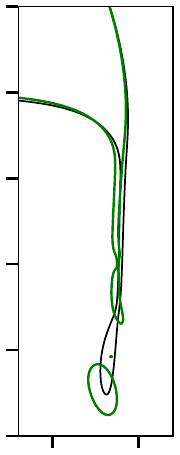}

\includegraphics{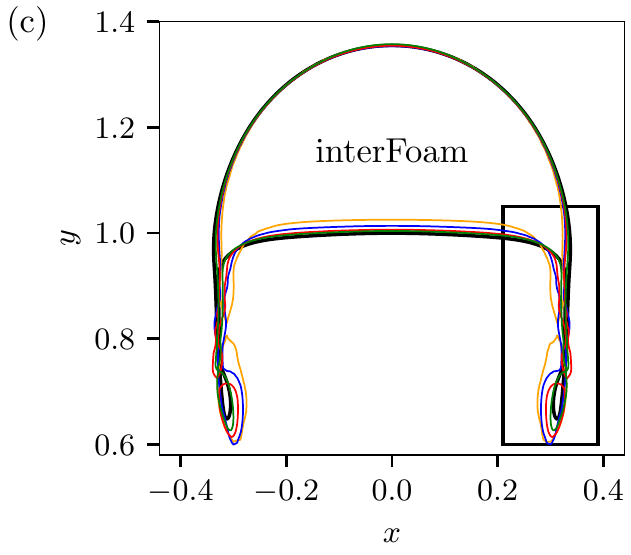}
\includegraphics{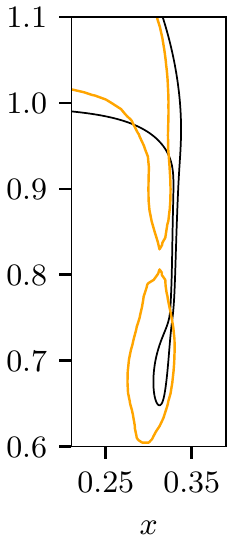}
\includegraphics{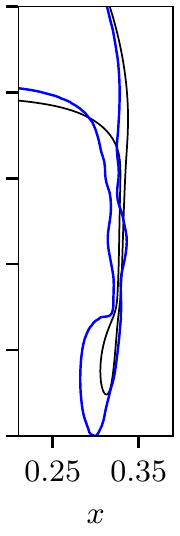}
\includegraphics{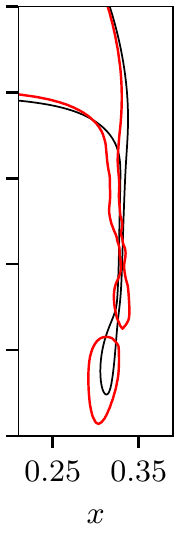}
\includegraphics{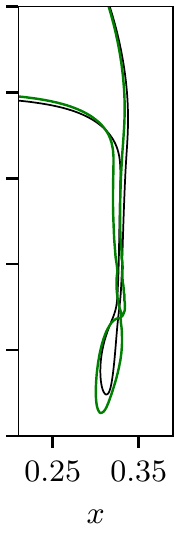}
\end{center}
\caption{\label{fig:risingInt1} Gas-liquid interface at $t=3$ s for 2D rising bubble problem using conventional schemes: (a) HF37. (b) RDF, and (c) interFoam. Left pane: the bubble on all considered resolutions and  right panes: convergence of the enlarged interface with grid refinement presented using consistent color coding with the left pane (shown in legend of (a)). }
\end{figure}

Figure~\ref{fig:risingInt1} presents the fluid-gas interface at time $t=3$ s for conventional methods. Solid black line is the reference solution. The bubble is significantly deformed at this point with a circular front pulling two slender trails, or tails, from backwards. The curvature is high at the end of the tails. On the coarsest grid, $N=80$, all conventional schemes fail to contain monolithic structure of the bubble. The tails broke up, and we observe large portions that are pinched off from the main body. 
HF37 method significantly improves its estimations with grid refinement. A large-scale break-up is not observed in the tails, and the bubble geometry is fairly well captured on finer grids. However, even on the finest grid ($N=640$), we observe a small-scale residue  of 10-15 cells that is broken from the tail, cf. the enlarged view of the tail in figure~\ref{fig:largeRise}a. At the tip of the tail the curvature is largest with a value around $\kappa_{\max}\approx 520 \mbox{ cm}^{-1}$ (not shown here). Given the grid spacing is $h=1/640$ cm, the non-dimensional curvature here is $\kappa_{max}h\approx 0.8$. While such a high curvature regime is also problematic for PLIC-based interface reconstruction, HF37 especially is known to have erroneous estimations in this range. The small-scale air residues are the manifestation of this limitation.

 \begin{figure}[!t]
\begin{center}
\includegraphics{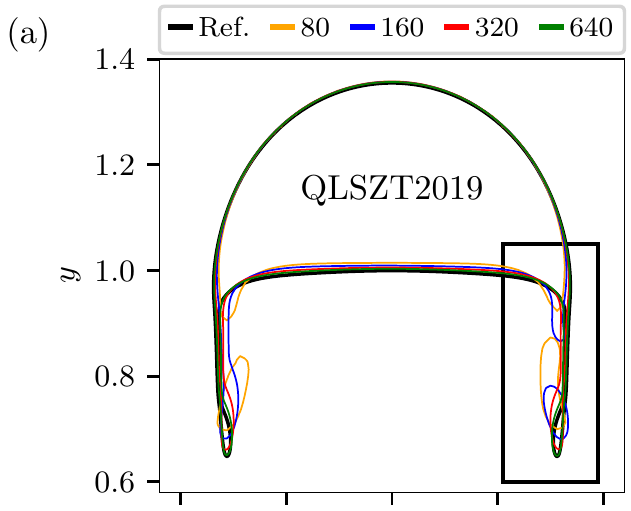}
\includegraphics{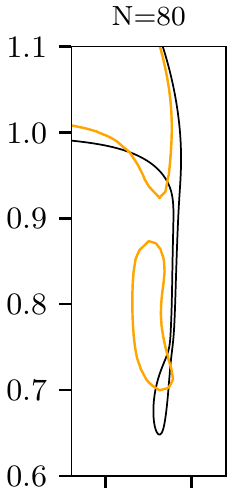}
\includegraphics{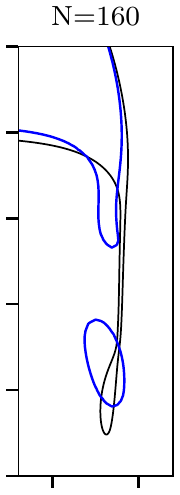}
\includegraphics{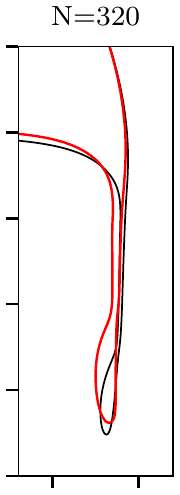}
\includegraphics{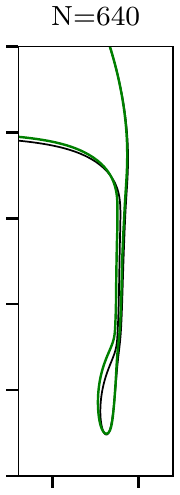}

\includegraphics{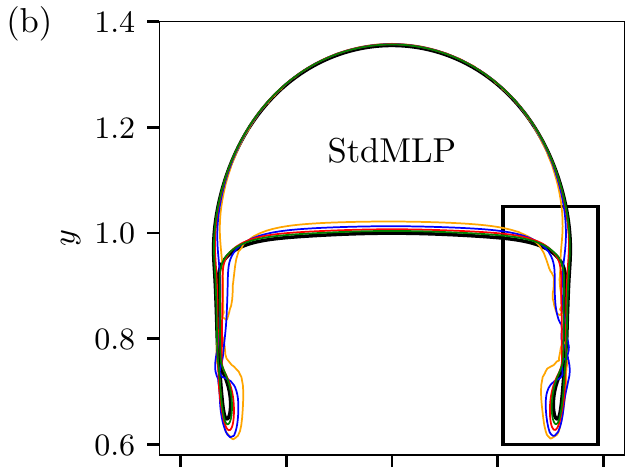}
\includegraphics{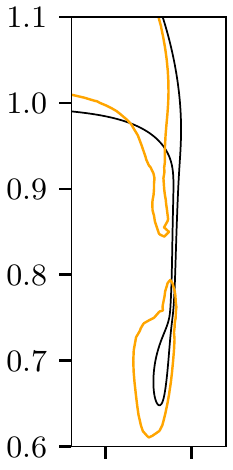}
\includegraphics{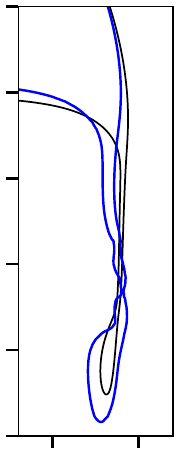}
\includegraphics{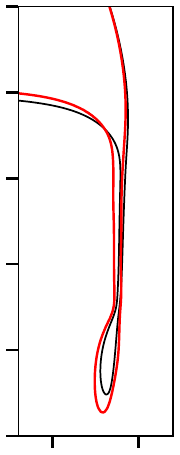}
\includegraphics{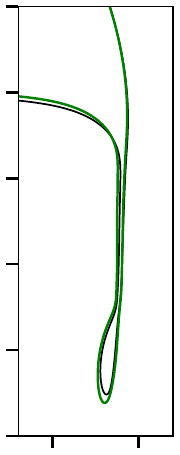}

\includegraphics{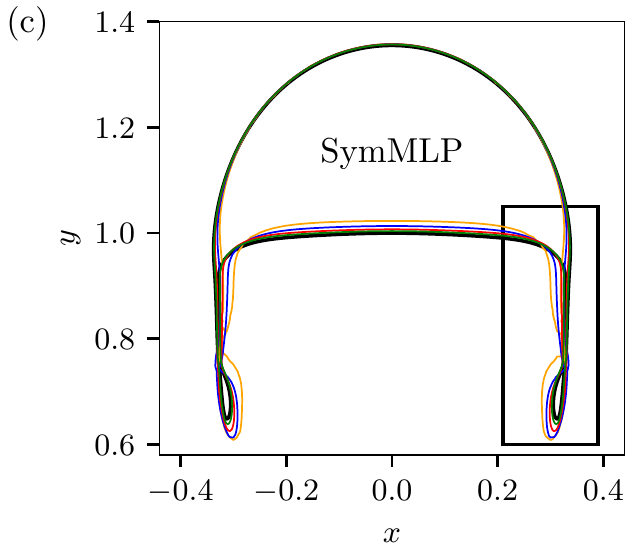}
\includegraphics{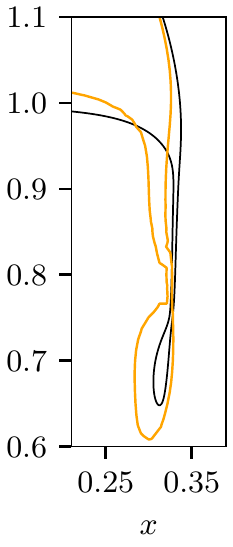}
\includegraphics{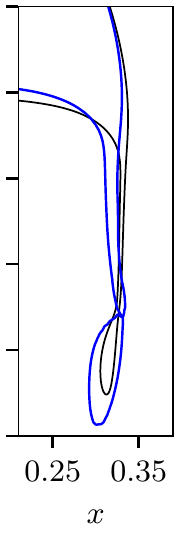}
\includegraphics{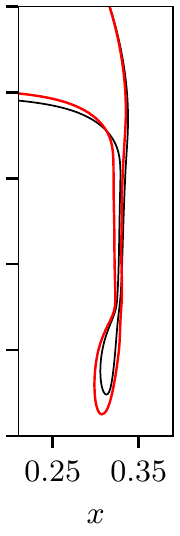}
\includegraphics{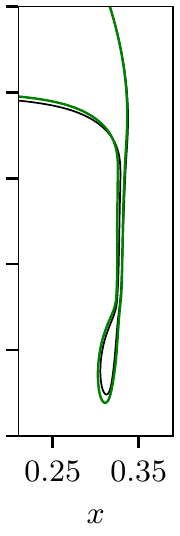}
\end{center}
\caption{\label{fig:risingInt2} Gas-liquid interface at $t=3$ s for 2D rising bubble problem using machine learning schemes. (a) QLSZT2019 \cite{Qi2019-mu}; (b) StdMLP; (c)~SymMLP. See the caption of figure~\ref{fig:risingInt1} for panes.} 
\end{figure}

Figure~\ref{fig:risingInt2} shows the bubbles obtained with MLP schemes.  Both standard MLP schemes exhibit bubble break-up on coarse resolution, cf. $N=80$ in figures~\ref{fig:risingInt2}a,b. QLSZT2019 further struggles on $N=160$ resolution, and  a large secondary bubble pinched off from the main body is observed. StdMLP manages to prevent bubble break-up on $N=160$ but has considerable wiggles at the tails. Symmetry-preserving MLP scheme delivers the most robust results. It is the only one out of six tested schemes that does not yield an artificial break-up in the tail on any of the resolutions, cf. figure~\ref{fig:risingInt2}c. On $N=160$ resolution, it has a smoother tail structure than StdMLP. All MLP schemes provide good estimations on finer grids. StdMLP and SymMLP deliver very similar results, whereas interfaces provided by QLSZT2019 have somewhat shorter tails but overall similar accuracy. The enlarged view on $N=640$ resolution in figures~\ref{fig:largeRise}b-d show that while QLSZT2019 predicts the tail end of the bubble very accurately, StdMLP and SymMLP predict the overall neck structure better. All MLP schemes outperform HF37 in the high-curvature region, as they don't leave any small-scale residue behind and provide interface geometries closer to the reference.

\begin{figure}[!t]
\begin{center}
\subfloat[HF37]{\includegraphics[scale=0.065]{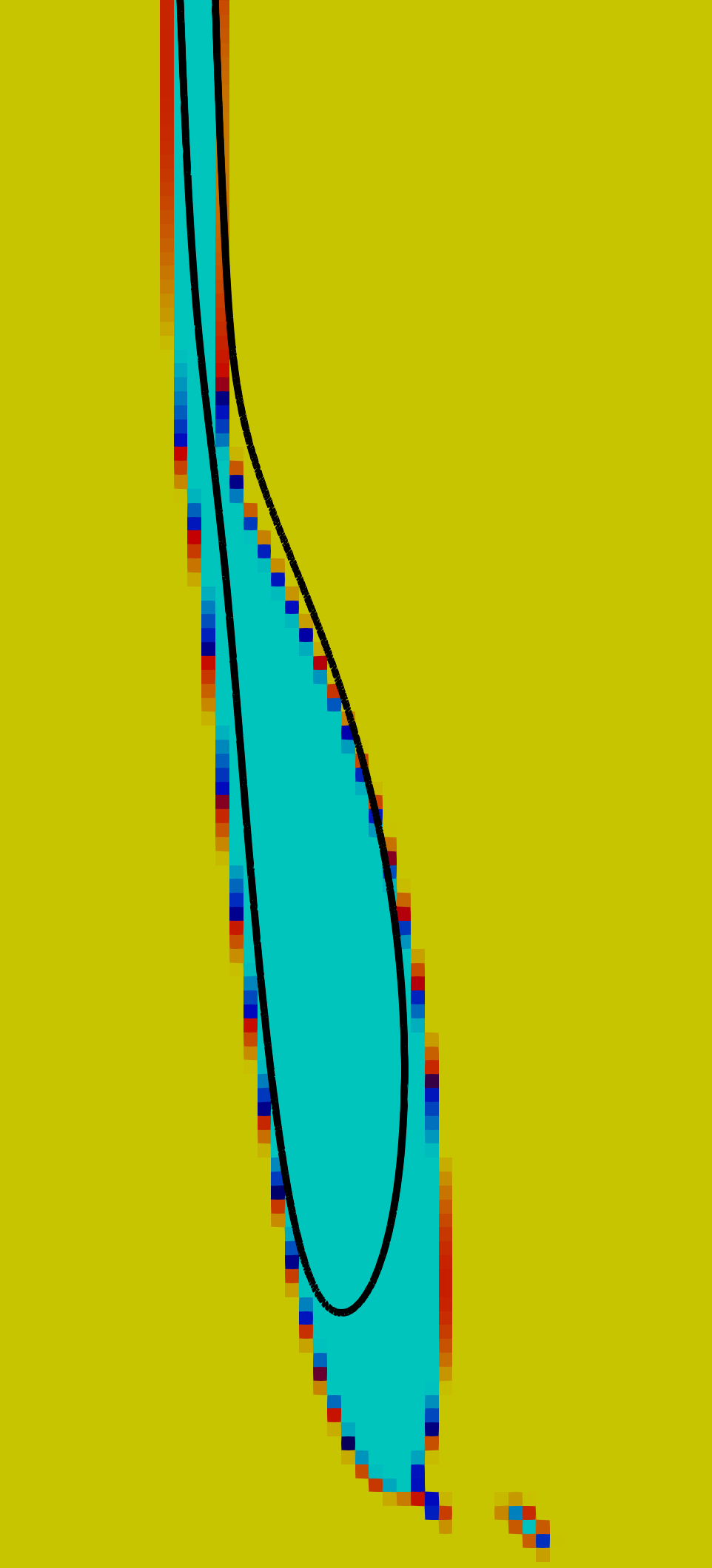}}
\subfloat[QLSZT2019]{~~~~~\includegraphics[scale=0.065]{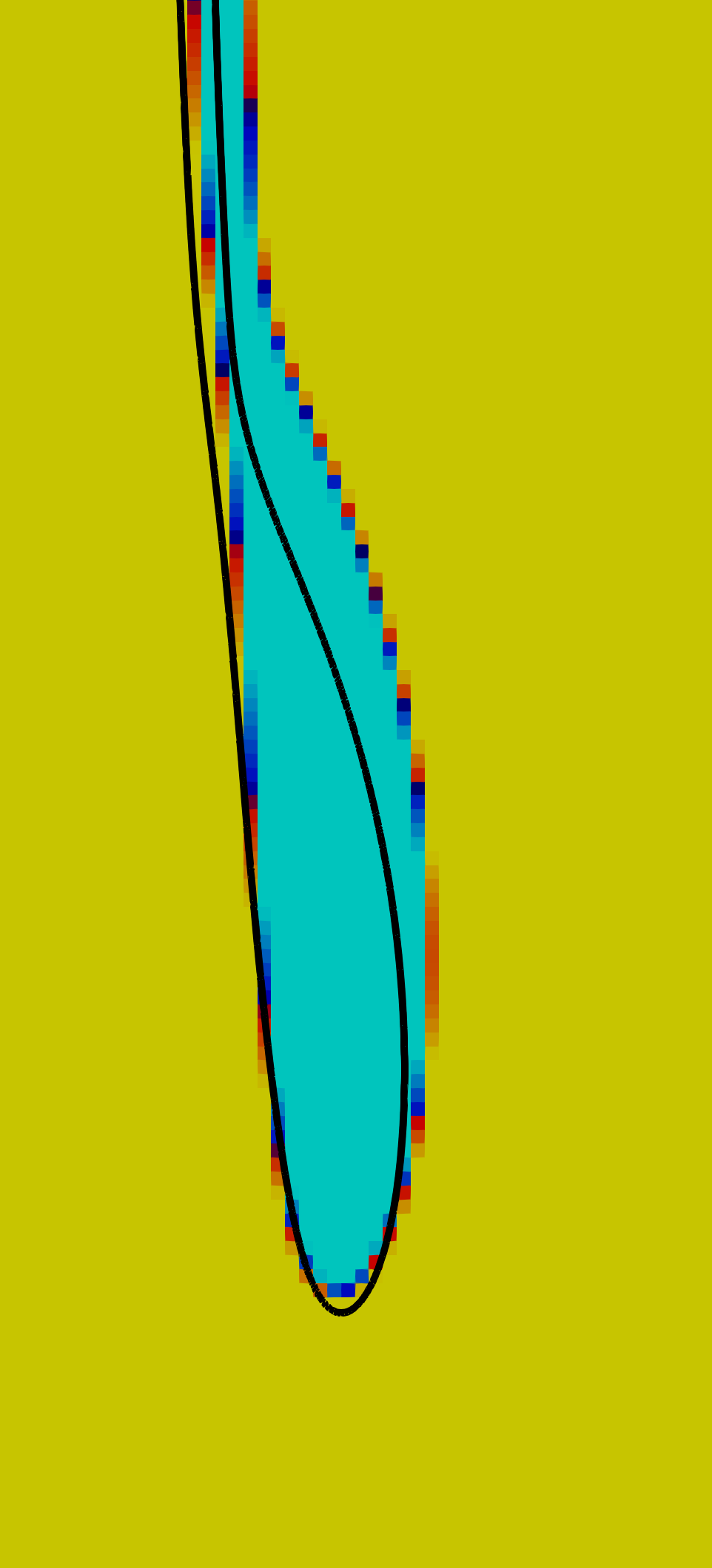}}
\subfloat[StdMLP]{~~~~~\includegraphics[scale=0.065]{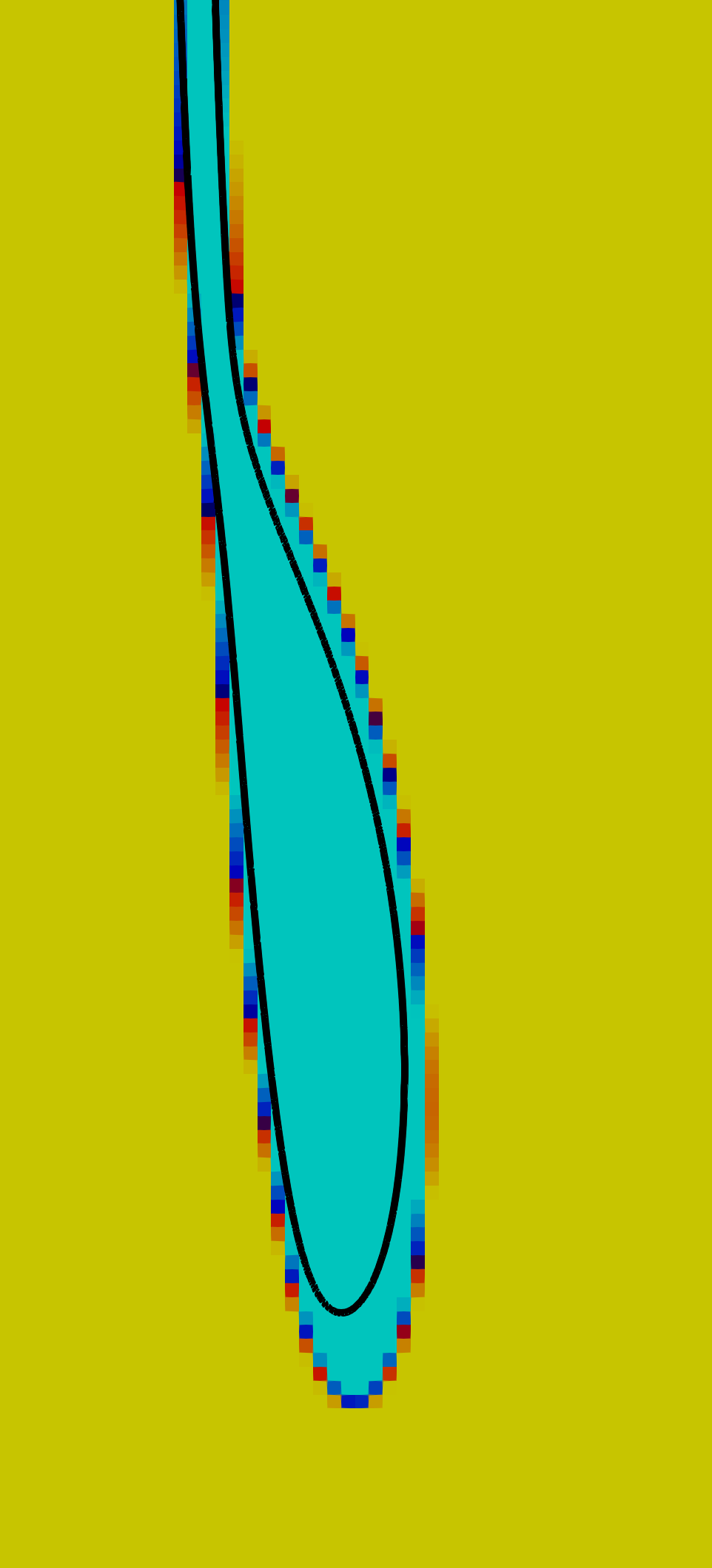}}
\subfloat[SymMLP]{~~~~~\includegraphics[scale=0.065]{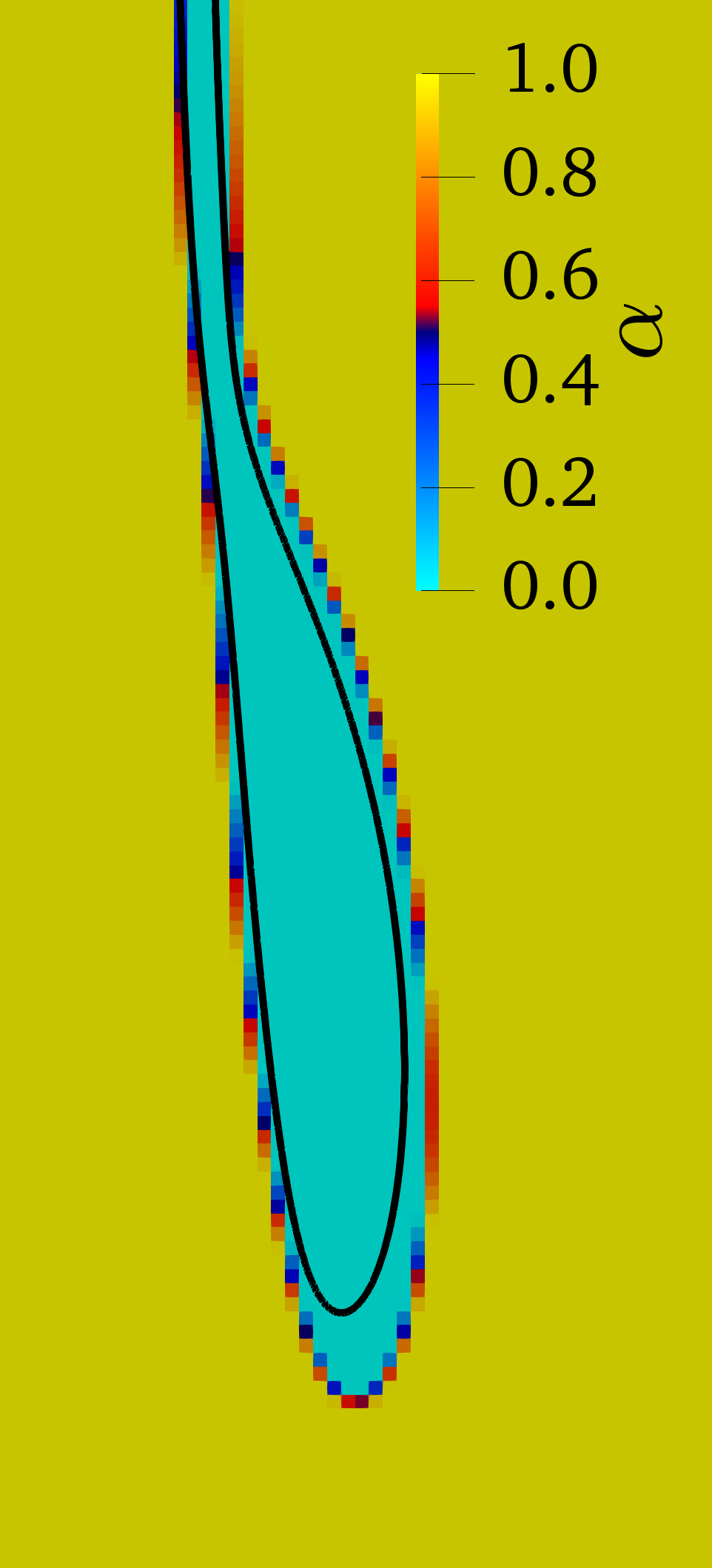}}
\caption{\label{fig:largeRise}  The left tail end of the deforming bubble with N=640 resolution. (a) HF37 (cf. figure~\ref{fig:risingInt1}a for the complete bubble); (b) QLSZT2019 (cf. figure~\ref{fig:risingInt2}a); (c) StdMLP (cf. figure~\ref{fig:risingInt2}b); (d) SymMLP (cf. figure~\ref{fig:risingInt2}c). Filled contours show cell-centred volume fractions data. Solid line represents the reference gas-liquid interface. }
\end{center}
\end{figure}

\subsection{Gravity-capillary wave with ripples}\label{sec:gcw}

The final test case is a two-dimensional nonlinear gravity-capillary wave that is unstable to parasitic ripples.  The dynamics of parasitic ripples are controlled by the interplay of capillary mechanisms and vorticity fields. Thus, numerical simulation of these small-scale features demand very accurate curvature models and fine computational grids resolving velocity gradients along and across the interface. Parasitic capillary ripples are well reproduced in single-phase flow simulations where the kinematic and dynamic free-surface conditions are explicitly satisfied, e.g., Refs. \cite{Mui1995-mr,Hung2009}. They are significantly more challenging to capture in VOF frameworks, and to our knowledge, there are no studies validating VOF method on these flows.  We select one of the simulations in Hung and Tsai (2009) \cite{Hung2009} as the test case. This configuration is a gravity capillary wave of $\lambda=5$~cm wavelength. The wave has initially the form of Fenton's fifth-order Stokes wave \cite{Fenton1985} with steepness $a_0k=0.25$ where $a_0$ is the half of the initial wave height and $k=2\pi/\lambda$ is the wavenumber. Hung and Tsai simulated this wave using a single-phase free-surface-flow code and obtained grid-converging results that will be used here for benchmarking.

\begin{figure}[!t]
\begin{center}
\includegraphics[scale=0.17]{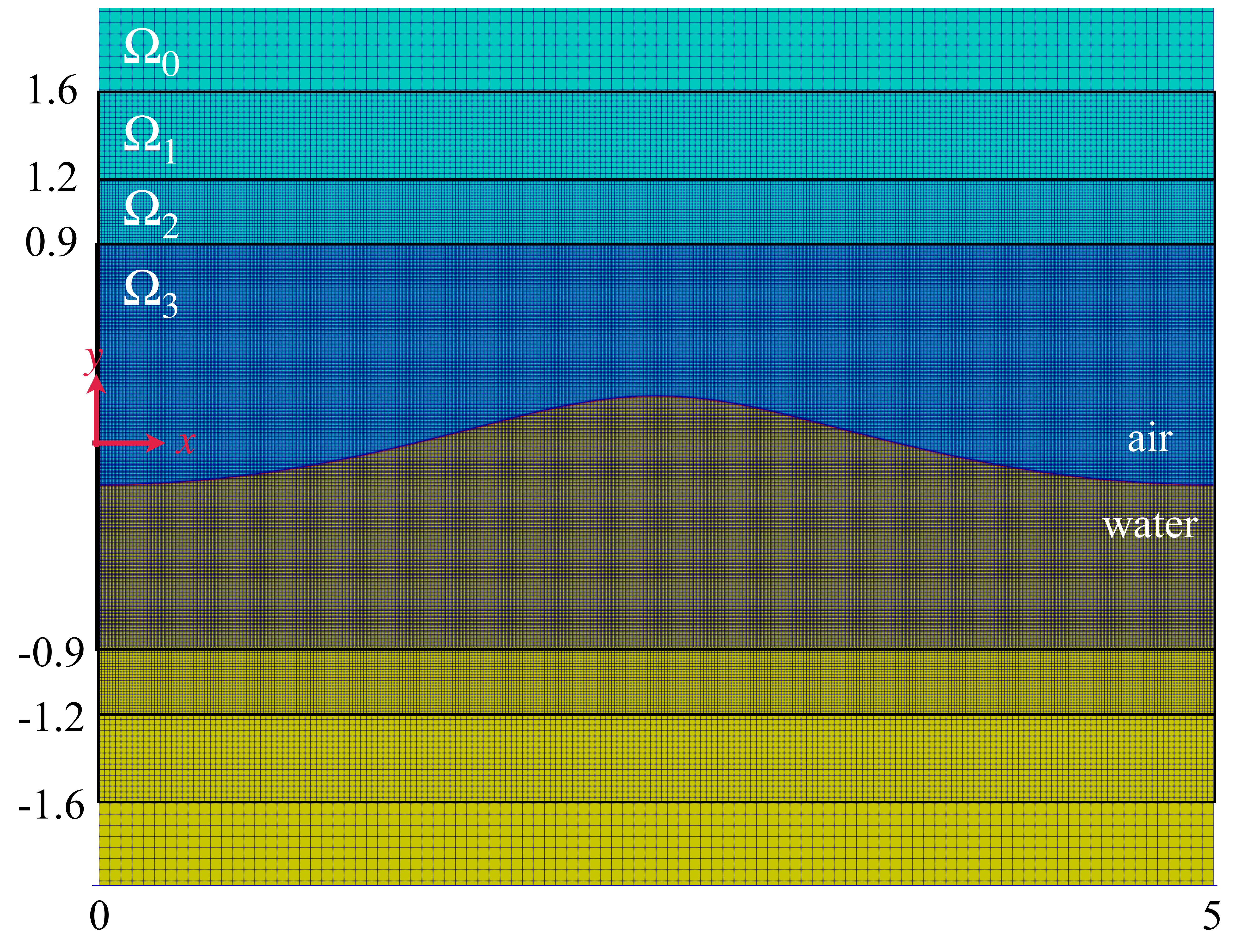}
\end{center}
\caption{\label{fig:tsaiGrid} Initial condition and the finest computational grid for gravity-capillary wave problem with three successive refinement zones $\Omega_1$, $\Omega_2$ and $\Omega_3$ applied on the base domain $\Omega_0:=[0, 5]\times[-7.5,7.5] \mbox{ cm}^2$ with resolution $100 \times 300$.   (yellow): $\alpha=1$; (cyan): $\alpha=0$.  }
\end{figure}

We consider a single wave in a horizontally periodic domain $\Omega_0:=[0,5]\times[-7.5,7.5] \mbox{ cm}^2$. The kinematic viscosity and density of the water are set to $\nu=10^{-6}\mbox{ m}^2\mbox{ s}^{-1}$ and $\rho=1000\mbox{ kg} \mbox{ m}^{-3}$. Unlike in Hung and Tsai (2009), our simulations contain the air phase with a kinematic viscosity and density of $\nu=1.48^{-5}\mbox{ m}^2\mbox{ s}^{-1}$ and $\rho=1.225\mbox{ kg} \mbox{ m}^{-3}$. A constant surface tension coefficient $\sigma=0.073$~kg~s$^{-2}$ is specified.  A Cartesian grid with a uniform resolution of $100 \times 300$ is constructed as the base grid. For finer grid levels, only the regions around the wave is locally refined.  Fig.~\ref{fig:tsaiGrid} presents the finest considered grid with three consecutive refinement zones $\Omega_1$, $\Omega_2$ and $\Omega_3$. In the finest zone $\Omega_3$, resolution per wavelength is $N=800$. A Courant number of $\mathrm{Co}=0.02$ is employed to restrict the maximum time step size.  

\begin{figure}[!t]
\begin{center}
\includegraphics[scale=1]{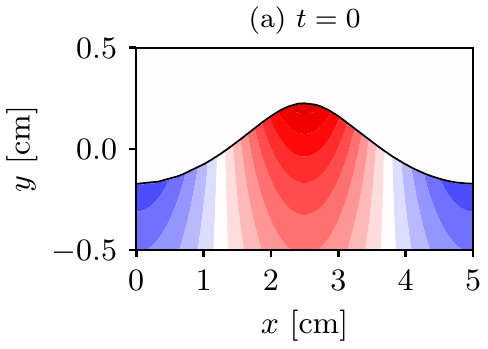}
\includegraphics[scale=1]{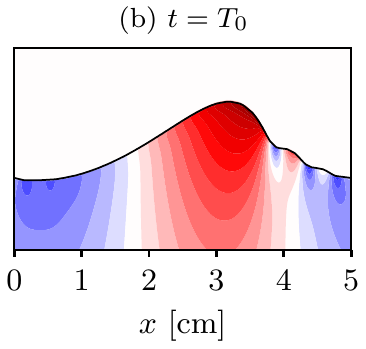}
\includegraphics[scale=1]{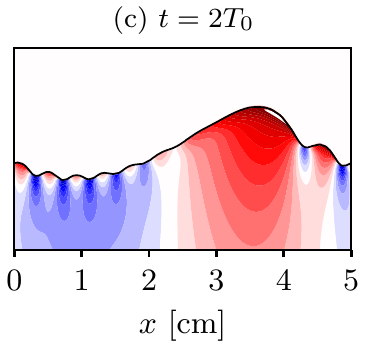}
\includegraphics[scale=1]{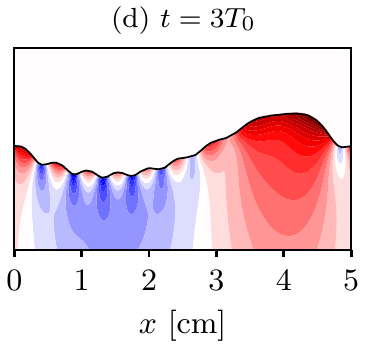}
\includegraphics[scale=1]{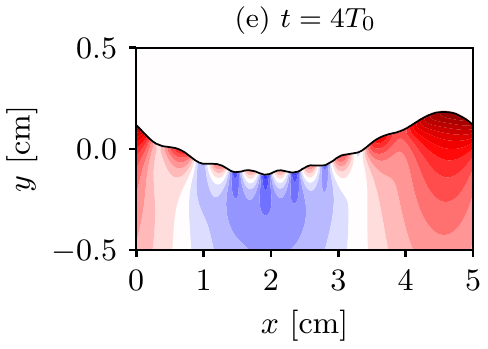}
\includegraphics[scale=1]{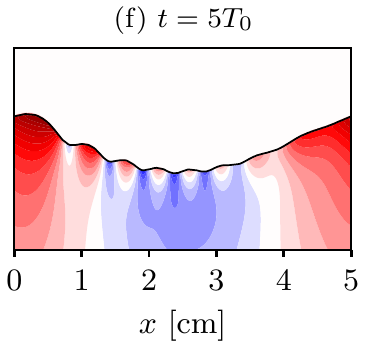}
\includegraphics[scale=1]{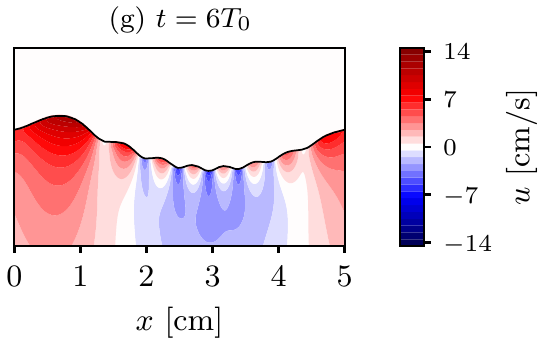}
\end{center}
\caption{\label{fig:6T_GCW} Horizontal velocity fields under the propagating gravity-capillary wave. The data is obtained with SymMLP on $N=800$ resolution. The wave propagates from left to right.}
\end{figure}

The linear period of the gravity-capillary wave is $T_0=2\pi(gk+\sigma k^3/\rho)^{-1/2}$ where $k=4\pi^2/(gT_0^2)$. The propagation of the wave and horizontal velocity fields are shown in figure~\ref{fig:6T_GCW} for 6 linear periods. The data in the figure is generated with SymMLP on $N=800$ resolution. The initial wave is symmetric with a steep crest and flatter trough. After one linear period, the crest leans forward and the first parasitic ripples develops at the forward face of the carrier wave. The ripples quickly spreads over the flatter portions of the carrier wave, and a train of parasitic ripples occupy the whole trough region, cf.~figure~\ref{fig:6T_GCW}b. The crests of small-scale ripples induce strong positive velocities beneath them. The wave reaches a quasi-equilibrium stage at later times, and there is no further onset of ripples. The reader is referred to Ref.~\cite{Hung2009} for a detailed discussion of the underlying physics.  

 \begin{figure}[!t]
\begin{center}
\includegraphics{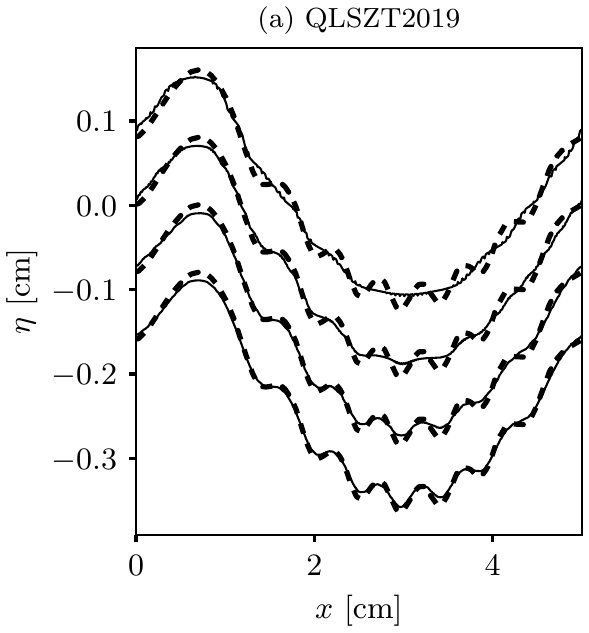}
~~\includegraphics{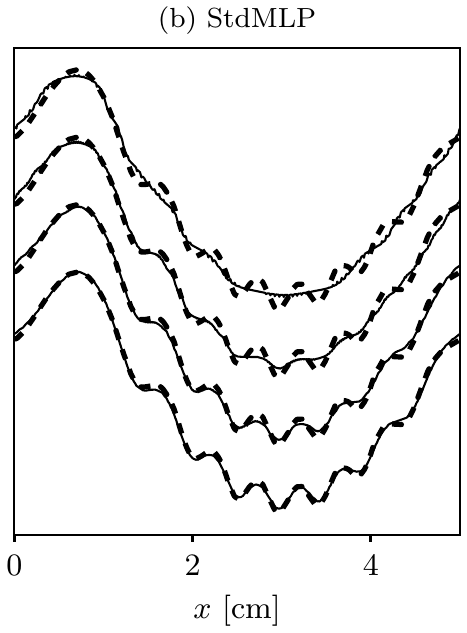}
~~\includegraphics{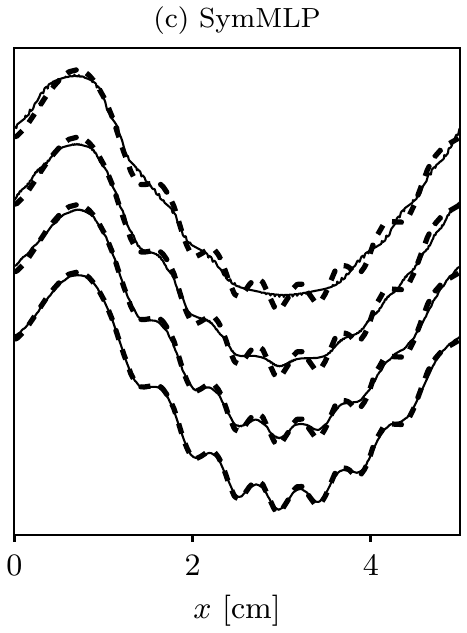}

\end{center}
\caption{\label{fig:etaTsai}. The water elevations at $t=6T_0$ on $N=100,200,400$ and $N=800$ resolutions (solid lines) are compared with the data from Hung and Tsai (2009) \cite{Hung2009} (dashed lines). Resolutions are per wavelength of the carrier wave $N=\lambda/h$. The data for each refined level is shifted down by $0.08$ cm. }
\end{figure}

Fig.~\ref{fig:etaTsai} compares the phase-adjusted water elevations at $t=6 T_0$ from MLP schemes to the data from Ref.~\cite{Hung2009}. All schemes fail to reproduce capillary ripples on coarse $N=100$ grid but capture well the height and the shape of the carrier wave. Models yield good estimations on finer grids capturing most of the characteristics of ripples. On the finest grid ($N=800$), all MLP schemes closely match the reference water elevation. 

\begin{figure}[!t]
\begin{center}
\includegraphics[]{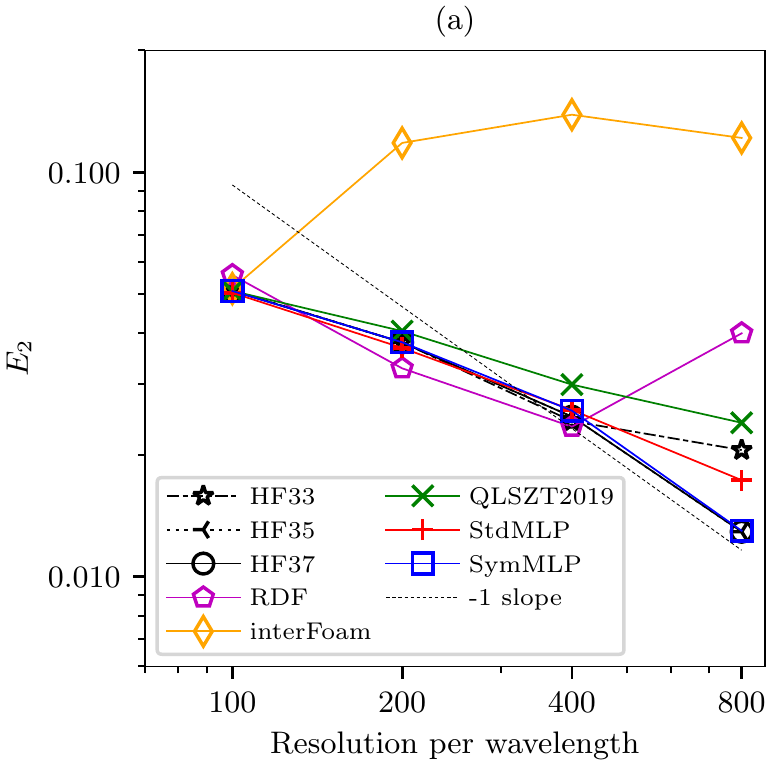}
~~~\includegraphics[]{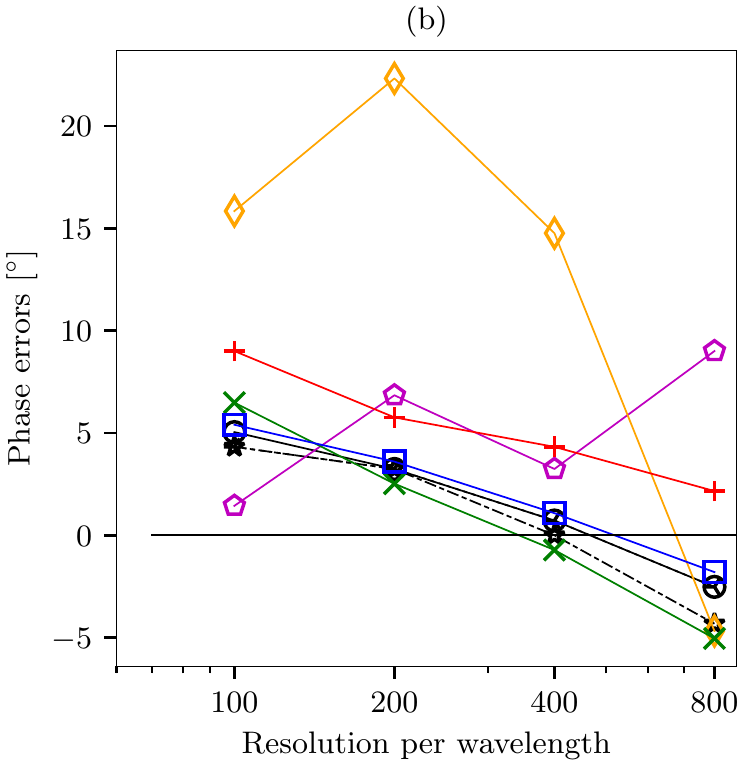}
\end{center}
\caption{\label{fig:ripplesErr} Convergence results for the parasitic ripples problem from Hung and Tsai (2009) \cite{Hung2009}. (a) Phase-adjusted water-elevation errors in $L_2$ norm (\ref{eq:E2Ripple}). (b)~Phase errors in degrees.}
\end{figure}

The discrepancies between the reference and simulated water elevations can be quantified by the following $L_2$ norm:  
\begin{equation}
    E_2(t)=\frac{1}{a}\sqrt{\frac{1}{\lambda}\int_0^{\lambda}(\eta(x,t)-\eta_{ref}(x,t))^2\mathrm d x},
    \label{eq:E2Ripple}
\end{equation}
where $\eta$ is the phase-adjusted water elevation. Figure~\ref{fig:ripplesErr}a presents the errors in water elevation. All three variants of the height function method, HF33, HF35 and HF37, are included in benchmarks. HF35, HF37 and SymMLP deliver similar results, and provide best overall estimations. Only in the finest-grid range these methods achieve close to first-order convergence. HF33 never approaches this convergence rate, and yields less accurate water elevation on the finest grid. 

So far, the wave is examined in phase-adjusted manner. Figure~\ref{fig:ripplesErr}b further plots the phase errors. HF35, HF37 and SymMLP exhibit smallest phase errors which range between $-2^\circ$ to $5^\circ$. The phase errors in these models converge to a slight negative value $-2^\circ$ with grid refinement instead of $0^\circ$. We note that the numerical simulations in Hung and Tsai (2009) were conducted using a single-phase model with only the water phase. Therefore, there was no air friction in their simulations, which might explain the slight phase lag in the converged results of HF35, HF37 and SymMLP.

\section{Conclusions} \label{sec:conclusions}
Data-driven models provide a promising alternative to estimate geometrical interface properties in the context of fixed-grid two-phase flow solvers. However, they are not sensitive to symmetries of the flow and can deliver unphysical results. The present work proposed a cost-effective methodology to preserve such symmetries in two dimensions on regular Cartesian grids. The methodology is based on feed-forward neural networks, or multilayer perceptrons (MLPs), that read volume fractions data in VOF-PLIC framework and return the non-dimensional curvature. MLP models are trained using synthetic datasets composed of circular arc segments. The anti-symmetries in the interface are preserved using input normalization, bias-free neurons and odd activation functions. Symmetries with respect to diagonal, vertical and horizontal reflections were further conserved by: (i) restricting the training to the regions where vertical normal dominates over the horizontal and allowing rotations; (ii) averaging over the remaining zones.  

We first tested the symmetry-preserving model on static configurations where the reference curvature can be obtained analytically. A standard MLP model is also developed and included in our tests along with the standard MLP model of Qi et al.~(2019) \cite{Qi2019-mu}. Furthermore, the height function method, a state-of-the art conventional method for curvature estimation on regular grids, is also considered in our benchmarks. The symmetry-preserving MLP model significantly outperforms both standard models, and delivers results that are mostly on par with the height function method. 

All curvature-estimation models are then implemented into the open-source CFD platform OpenFOAM and validated on several standard benchmark problems. The symmetry-preserving MLP model effectively reduces the spurious currents around a stationary bubble to the levels of machine precision or close. Standard MLP models suffer from lack of symmetries and exhibit high-intensity spurious currents when the full stationary circle is considered. The new model also delivers better  estimations on other surface-tension dominated flows such as standing and parasitic capillary waves and a rising bubble. The results of the symmetry-preserving model are competitive with those of the height function method using $3\times 7$ stencil with the former even surpassing the latter in high-curvature regions. This is despite the fact that the symmetry-preserving MLP using a smaller $3\times 3$ stencil. On $3\times 3$ stencil, the height function method delivers vastly inferior results than its $3\times 7$ counterpart and the symmetry-preserving MLP.  

{\color{black}The present methodology is designed for two-dimensional interfaces. A straightforward extension to three-dimensions can be done using datasets composed of spherical segments. However, in three dimensions, the mean curvature has two principal components, which are always identical in a spherical interface. Therefore, spherical interface segments do not provide a comprehensive dataset. They cannot capture saddle points where one principal component is positive and another negative. Training datasets must be extended to more complex interface configurations such as two-dimensional waves, significantly increasing the training costs. In this regard, our methodology can provide even more drastic benefits in three dimensions. To this end, odd-symmetric activation functions can readily be adapted and reduce the dimension of the dataset by 50\%. Additionally, restricting training to the configurations where a selected component of the normal vector dominates and allowing rotations will further reduce the cost by 66\%. Consequently,  training and evaluation of the MLP models can be done in 1/6th of all possible configurations increasing the feasibility of averaging operation to preserve symmetries. }

Machine learning models, test cases and C++ code for OpenFOAM implementations can be found in the git repository: \url{https://github.com/asimonder/geometricVoFCartesian}.

\section*{Acknowledgements}
P. L.-F. Liu would like to acknowledge the supports from the Ministry of Education in Singapore through a research grant (MOE2018-T2-2-040). This research was also supported in part by Yushan Program, Ministry of Education in Taiwan. The  computational  work for this article was fully performed on resources of the National Supercomputing Centre, Singapore (\url{https://www.nscc.sg}). 

\appendix 

\section{Implementation and solver details}
Open-source CFD code OpenFOAM v2006 \cite{ofv2006} is selected as the simulation platform.  The implementation details of MLP schemes are briefly surveyed here. OpenFOAM employs a finite volume method designed for generic unstructured grids. It is based on a collocated approach, where all solution variables, hence the volume fraction data, are stored in cell centroids. Being a generic solver, OpenFOAM does not support Cartesian \textit{ijk} addressing of regular grids. Therefore, in the first step, we have implemented a new class \textit{ijkZone} that allows access to arbitrarily many neighbours of the cell using $ijk$ addressing. The class works on a subset of the computational mesh where the grid is uniform and the fluid-fluid interface is located. Utilizing \textit{ijkZone}, we can access the volume fraction data in the block surrounding each cut cell. In the second step,  \textit{multilayerPerceptron} class is implemented, which covers basic functionalities and data structures of an MLP. The class reads the volume fractions in the cell stencil and returns the model output. For optimum performance, data structures encompassing MLP are implemented using the \textit{Standard C++ Library}. To this end, weights and biases are stored in nested data structures \textit{std::vector$<$std::vector$<$std::vector$<$double$>>>$} and \textit{std::vector$<$std::vector$<$double$>>$}, respectively. These dynamic containers provide the flexibility required to load models of varying complexity on runtime.  Finally, classes concerning MLP models, named \textit{mlpCurvature} are implemented at a higher level. For benchmarking, the height function method is implemented into OpenFOAM library as well. This method also requires uniform square cells. Thus, \textit{ijkZone} class is utilized in its implementation.

\begin{figure}[!t]
\begin{center}
\includegraphics[scale=0.29]{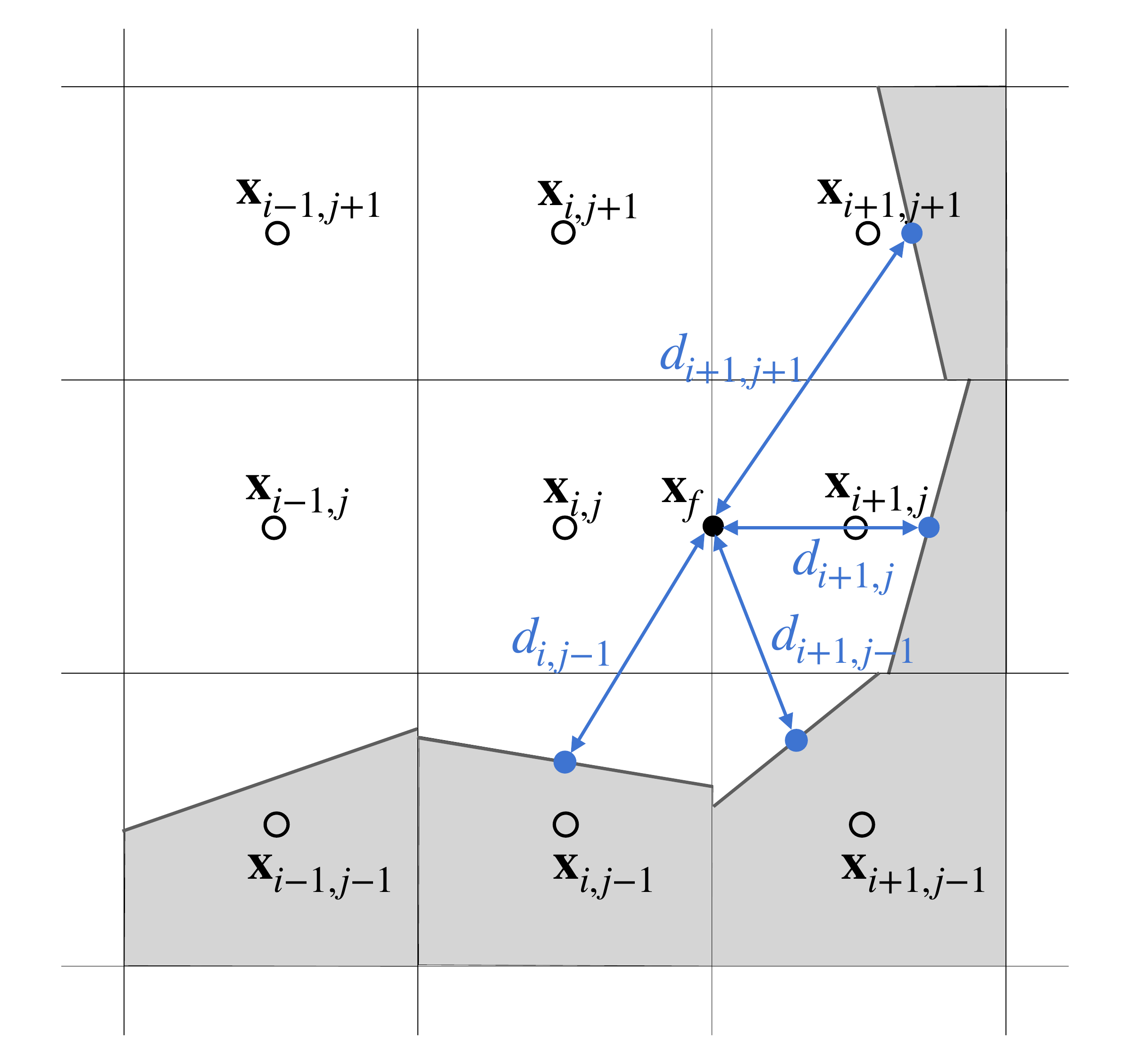}
\caption{\label{fig:KappaInt} Distance measures for inverse-distance weighted extrapolation to a face between a cut and non-cut cell.}
\end{center}
\end{figure}

In OpenFOAM, the curvature is evaluated at face centres to enforce a discrete balance between pressure and surface tension forces in stationary conditions. The face-centred value of curvature between two cut cells is found by linearly interpolating two neighbouring cell-centred values.  The linear interpolation is ineffective when the face is located between a cut cell and non-cut cell, as the curvature is set to zero in non-cut cells.  For these faces, the curvature is extrapolated from neighbouring cells. To that end, we employ a kernel that is based on inverse-distance weights, i.e.,
\begin{equation}
\kappa_f=
\begin{cases}
  \frac{1}{2}\left (\kappa_P+\kappa_N \right) & \text{if $\kappa_p\neq 0$ and $\kappa_N \neq 0$,} \\
  \frac{\sum_i w_i \kappa_i}{\sum_i w_i}& \text{if $\kappa_p \neq 0$ and $\kappa_N=0$,} \\
  0 & \text{otherwise,}
\end{cases}
\label{eq:intFace}
\end{equation}
where $\kappa_P$ and $\kappa_N$ are curvature values in the neighbouring cells of the face, the subscript $i$ represents the cells in the local cell block around the face, and $w_i$ is the interpolation weights given by
\begin{equation}
w_i=
\begin{cases}
  \frac{1}{d_i^p} & \text{if $d_i <3/2 h$ and $\kappa_i>0$} \\
  0 & \text{otherwise}
\end{cases}
\end{equation}
with $d_i$ being the distance between the center of the reconstructed interface segment and the center of the face, cf. figure~\ref{fig:KappaInt}. We set the power parameter to $p=2$. The interpolation scheme  (\ref{eq:intFace}) is employed in machine learning models and height function method.

\bibliographystyle{elsarticle-num}
\bibliography{allpapers.bib}

\end{document}